\documentclass[english,a4paper,headsepline]{article}
\usepackage[cp850]{inputenc}    
\usepackage{babel}
\usepackage{amsmath}
\usepackage{epsfig}
\newcommand{\E}{\ensuremath{\mathrm{e}}}

\newcommand{\bra}[1]{\ensuremath{\langle #1|}}
\newcommand{\ket}[1]{\ensuremath{|#1\rangle}}

\newcommand{\braket}[2]{\ensuremath{\langle #1|#2\rangle}}

\begin{document}

\title{Bayesian Approach to Inverse Quantum Statistics:
Reconstruction of Potentials in the Feynman Path Integral
Representation of Quantum Theory}

\author{J. C. Lemm, J. Uhlig, and A. Weiguny
       \\{\small Institut f\"ur Theoretische Physik, 
          Universit\"at M\"unster}}

\maketitle

\begin{abstract}
The Feynman path integral representation of quantum theory is used
in a non--parametric Bayesian approach to determine quantum potentials from
measurements on a canonical ensemble.
This representation allows to study explicitly the classical and semiclassical
limits and provides a unified description in terms of functional integrals:
the Feynman path integral for the statistical operator,
and the integration over the space of potentials  
for calculating the predictive density.
The latter is treated in maximum a posteriori approximation, and various
approximation schemes for the former are developed and discussed.
A simple numerical example shows the applicability of the method.

\end{abstract}


\renewcommand{\thesection}{1}
\section{Introduction}

The solution of the quantum many--body problem requires both techniques for
solving the Schr\"odinger equation and knowledge of the underlying forces,
often to be deduced from observational data.
In the field of nuclear physics, forces are extracted from scattering data 
and ground state properties of the two--nucleon system, 
since no practicable basic theory of nuclear forces exists up to date.
Given a data--based phenomenological nucleon--nucleon potential, one can, 
in principle, construct the related potential between two colliding nuclei.
However, this is a formidable task which has been attacked only for a
few simple cases in an approximate way, and a direct calculation of the 
nucleus--nucleus potential from observational data is highly desirable 
for practical applications like, e.g., in nuclear astrophysics.
In solid state physics the basic force is known to be the Coulomb force,
however, for a straight problem like the motion of a single electron 
under the influence of a crystal surface, one would prefer to deduce the 
respective potential directly from observational data rather than going 
through the full many--body problem of electrons and nuclei of the crystal.

The reconstruction of such two--body forces or single--particle potentials
from experimental data constitutes a typical inverse problem of quantum theory.
Such problems are notoriously ill--posed in the sense of Hadamard \cite{1}
and require additional a priori information to obtain a unique, stable 
solution. Well--knowm examples are inverse scattering theory \cite{2}
and inverse spectral theory \cite{3}. They describe the kind of data 
which are necessary, in addition to a given spectrum, to identify the
potential uniquely. For example, such data can be a second spectrum
for different boundary conditions, knowledge of the potential on a 
half--interval or the phase shifts as a function of energy.
However, neither a complete spectrum nor specific values 
of the potential or phase shifts at all energies can be
inferred by a finite number of measurements.
Hence any practical algorithm for extracting two--body forces
or single--particle potentials from experimental data must rely 
on additional a priori assumptions like symmetries, smoothness, 
or asymptotic behaviour.
If the available data refer to a system at finite temperature
$T\ne 0$, one is led to the inverse problem of quantum statistics.
In such a case, non--parametric Bayesian statistics \cite{4}
is especially well suited to include both observational data and
a priori information in a flexible way.

In a series of papers \cite{5}, the Bayesian approach 
to inverse quantum statistics has been applied to reconstruct potentials
(or two--body interactions) 
from particle--position measurements on a canonical ensemble.
A priori information was imposed through approximate symmetries (translational,
periodic) or smoothness of the potential or 
by fixing the mean energy of the system.
The likelihood model of quantum statistics
(defining the probability for 
finding the particle at some position $x$ for a system with potential $V$
at temperature $T$)
was treated in energy representation.

In the present paper we apply the Feynman path integral representation
of quantum mechanics \cite{6} to calculate the statistical operator $\rho$
and related quantities in coordinate space. 
This representation is of interest in the context of inverse problems
in Bayesian statistics for two reasons:
First, it allows to study the transition to the semiclassical
and classical limits,
relevant for example to atomic force microscopy \cite{Fuchs}
so far treated on the level of classical mechanics.
However, scales may soon be reached where the inclusion of quantum effects
will be mandatory. 
Second, one obtains a unified description of Bayesian statistics
in terms of path integrals.
These are on one side the Feynman path integrals, needed in the likelihood
model, and on the other side the functional integral over the space of 
potential functions $V$ when calculating the predictive density as integral
over the product of likelihood and posterior for all possible potentials.

Our paper is organized as follows:
An introduction to Bayesian statistics is presented in section 2,
showing how Bayes' theorem about the decomposition of joint probabilities 
can be used for the inverse problem of quantum statistics.
A general expression is given for the likelihood of a quantum system
with given potential $V$ for a canonical ensemble,
and the prior density is chosen as Gaussian process to implement
a bias towards smoothness and/or periodicity of the potential $V$.
This potential can be calculated from a non--linear differential equation
which results from the maximum posterior approximation for the predictive 
density.
Two approximation schemes for solving the inverse problem of quantum 
statistics in path integral representation are developed.
In the first variant (sections 3 and 4),
the path integral in the likelihood is treated in stationary phase 
approximation.
The resulting stationarity equations with respect to the path are classical
equations of motion for a particle in potential $-V$.
These equations are to be solved simultaneously with the stationarity
equations with respect to the potential, following from the maximum posterior
approximation in the above classical approximation of the path integral.
In the second variant (section 5), the basic stationarity equations
of the maximum posterior approximation are treated in terms of the Feynman
integrals. These equations which involve the logarithmic derivatives
of the statistical operator and the partition function are still exact 
in the sense of quantum theory.
Approximation schemes for the statistical operator in coordinate representation
and the corresponding partition function and their derivatives are developed
in sections 6 and 7.
In section 6, the quadratic fluctuations for the statistical operator
and the partition function around the classical paths of section 4 
are determined, while section 7 deals with the respective derivatives.
Three approximation schemes are studied under the general strategy
that the statistical operator drops out in the logarithmic derivative.
A simple numerical example is added in section 8 to show that the path
integral formalism can actually be used for problems of inverse quantum
statistics.

\renewcommand{\thesection}{2}
\section{Bayesian Approach to Inverse Quantum Statistics}
The aim of this paper is to determine the dynamical laws of quantum systems
from measurements on a canonical ensemble. The method used is
non-parametric
Bayesian inference combined with the path integral representation of
quantum
theory which allows to study the transition to the classical limit. To be
specific,
we aim at reconstructing the potential $V$ of the system from measurements
of the
position coordinate $\hat{x}$ of the particle for a canonical ensemble at
temperature $1\,/\,\beta$.

The general Bayesian approach, 
tailored to the above problem, 
is based on two probability densities:
\begin{enumerate}
\item
a likelihood $p\,(x|O,\,V)$ for the probability of outcome $x$ when
measuring observable $O$
for given potential $V$, not directly observable, and
\item
a prior density $p\,(V)$ defined on the space ${\cal{V}}$ of possible
potentials $V$.
\end{enumerate}
This prior gives the probability for $V$ before data have been collected.
Hence it has to comprise
all a priori information available for the potential, like symmetries or
smoothness. The need for a prior model, complementing the likelihood model,
is characteristic
for empirical learning problems which try to deduce a general law from
observational data.

These ingredients, likelihood and prior, are combined in Bayes theorem to
define the
posterior of $V$ for given data $D$ through
\renewcommand{\theequation}{2.1}
\begin{equation}
p\,(V|D)\,=\,\frac{p\,(x_{T}|O_{T},\,V)\,p\,(V)}{p\,(x_{T}|O_{T})}\,\,.
\end{equation}
Eq. (2.1) is a direct consequence of the decomposition of the joint
probability
$p\,(A,\,B)$ for two events $A,\,B$ into conditional probabilities
$p\,(A|B)$ and $p\,(B|A)$, respectively.
Observational data $D$ are assumed to consist of $N$ pairs,
\renewcommand{\theequation}{2.2}
\begin{equation}
D = \{(x_{i},\,O_{i})|1\,\leq\,i\,\leq\,N\}\,\,,
\end{equation}
where $x_{T}$, $O_{T}$ denote formal vectors with components $x_{i}$,
$O_{i}$.
Such data are also called training data, hence the label $_{T}$. For
independent data
the likelihood factorizes as
\renewcommand{\theequation}{2.3}
\begin{equation}
p\,(x_{T}|O_{T},\,V)\,=\,\prod\limits_{i}\,p\,(x_{i}|O_{i},\,V)\,\,,
\end{equation}
where a chosen observable $O_{i}$ may be measured repeatedly to give values
$x_{i}$,
equal or different among each other. The denominator in (2.1) can be viewed
as normalization factor
and can be calculated from likelihood and prior by integration over $V$,
\renewcommand{\theequation}{2.4}
\begin{equation}
p\,(x_{T}|O_{T})\,=\,\int\,DV\ p\,(x_{T}|O_{T},\,V)\,p\,(V)\,\,.
\end{equation}
The $V$-integral in eq. (2.4) stands for an integral over parameters, if we
choose a parametrized space ${\cal{V}}$ of potentials, or for a functional
integral over
an infinite function space.

To predict the results of future measurements on the basis of a data set $D
$, one
calculates according to the rules of probability theory the predictive
density
\renewcommand{\theequation}{2.5}
\begin{equation}
p\,(x|O,\,D)\,=\,\int DV\ p\,(x|O,\,V)\,p\,(V|D)
\end{equation}
which is the probability of finding value $x$ when measuring observable $O
$ under the
condition that data $D$ are given.
Here we have assumed that the probability of $x$ is completely
determined by giving potential $V$ and observable $O$, and does not depend
on training data
$D$, $p\,(x|O,\,V,\,D)\,=\,p\,(x|O,\,V)$, and that the probability for
potential $V$
given the training data $D$ does not depend on observable $O$ selected in
the future, $p\,(V|O,\,D)\,=\,
p\,(V|D)$.

The integral (2.5) is high-dimensional in general and difficult to
calculate in practice.
Two approximations are common
in Bayesian statistics: The first one is an evaluation of the integral by
Monte Carlo
technique. The second one, which we will pursue in this paper, is the so
called
maximum a posteriori approximation. Assuming the posterior to be
sufficiently
peaked around its maximum at potential $V^{*}$, the integral (2.5) is
approximated by
\renewcommand{\theequation}{2.6}
\begin{equation}
p\,(x|O,\,D)\,\approx\,p\,(x|O,\,V^{*})
\end{equation}
where
\renewcommand{\theequation}{2.7}
\begin{equation}
V^{*}\,=\,\text{argmax}_{V\,\in\,{\cal{V}}}\,p\,(V|D)\,=\,
\text{argmax}_{V\,\in\,{\cal{V}}}\,p\,(x_{T}|O_{T},\,V)\,
p\,(V)
\end{equation}
according to eq. (2.1) with the denominator independent of $V$. Maximizing
the posterior $p\,(V|D)$ with
respect to $V\,\in\,{\cal{V}}$ leads to solving the stationarity equations
\renewcommand{\theequation}{2.8}
\begin{equation}
\delta_{V}\,p\,(V|D)\,=\,0\,=\,\delta_{V}\,(p\,(x_{T}|O_{T},\,V)\,p\,(V))
\end{equation}
where $\delta_{V}$ denotes the functional derivative $\delta\,/\,\delta\,V
$. Equivalent to (2.8)
and technically often more convenient is the condition for the
log-posterior
\renewcommand{\theequation}{2.9}
\begin{equation}
\delta_{V}\,\ln\,p\,(V|D)\,=\,0\,=\,\delta_{V}\,\ln\,p\,(x_{T}|O_{T},\,V)\,
+\,
\delta_{V}\,\ln\,p\,(V)
\end{equation}
which minimizes the energy $E\,(V|D)\,=\,-\ln\,p\,(V|D)$ and will be used
in the following.

A convenient choice for prior $p\,(V)$ is a Gaussian process,
\renewcommand{\theequation}{2.10}
\begin{equation}
p\,(V)\,\sim\,\exp\,\left\{-\frac{\gamma}{2}\,\braket{V\,-\,V_{0}|K}{V\,
-\,V_{0}}\right\}\,=\,
\E^{-\frac{\gamma}{2}\,\Gamma\,[v]}
\end{equation}
where
\renewcommand{\theequation}{2.11}
\begin{equation}
\Gamma\,[v]=\braket{V-V_{0}|K}{V-V_{0}}=\int dx\,dx^{\prime}\,
\left[v\,(x)-v_{0}\,(x)\right]\,K\,(x,\,x^{\prime})\,[v\,(x^{\prime})-v_{0}\,(x^{\prime})],
\end{equation}
assuming a local potential $V\,(x,\,x^{\prime})\,=\,v\,(x)\,\delta\,(x\,
-\,x^{\prime})$. The mean $V_{0}$ represents
a reference potential or template for $V$, and the real-symmetric, positive
(semi-)definite covariance operator
$(\gamma\,K)^{-1}$ acts on the potential, measuring the distance between $V
$ and $V_{0}$. The hyperparameter $\gamma$ is used to
balance the prior against the likelihood term and is often treated in
maximum a posteriori approximation or determined 
by cross--validation techniques.   
A bias towards smooth functions $v\,(x)$ can be implemented by $K\,=\,
-d^{2}\,/\,dx^{2}$ choosing
$v_{0}\,(x)\,\equiv\,0$. If some approximate symmetry of $v\,(x)$ is
expected, like for a
surface of a crystal deviating from exact periodicity due to point defects,
one may implement a non-zero
periodic reference potential $v_{0}\,(x)$ in eq. (2.11).

The likelihood for our problem follows from the axioms of quantum theory:
The
probability to find value $x$ when measuring observable $O$ for a quantum
system in a state
described by a statistical operator $\rho\,=\,\rho\,(V)$ is given by
\renewcommand{\theequation}{2.12}
\begin{equation}
p\,(x|O,\,V)\,=\,{\rm Tr}\,\{P_{O}\,(x)\,\rho\,(V)\}
\end{equation}
where $P_{O}\,(x)\,=\,\Sigma_{\xi}\,\ket{x,\,\xi}\,\bra{x,\,\xi}$ projects
on the space spanned by
the orthonormalized eigenstates $\ket{x,\,\xi}$ of operator $O$ with
eigenvalue $x$, and the label
$\xi$ distinguishes degenerate eigenstates with respect to $O$. If the
system is not prepared
in an eigenstate of observable $O$, a quantum mechanical measurement will
change the state
of the system, i.\,e., will change $\rho$. Hence to perform repeated
measurements under same
$\rho$ requires the restoration of $\rho$ before each measurement. For
canonical ensembles at
given temperature,
\renewcommand{\theequation}{2.13}
\begin{equation}
\rho = \frac{\exp(-\beta\,H)}{{\rm Tr}\exp(-\beta\,H)}
\end{equation}
with Hamiltonian $H\,=\,T\,+\,V$ and temperature $1\,/\,\beta$, this means
to wait between two
consecutive observations until the system is thermalized again. Choosing
the particle
position operator $\hat{x}$ as observable $O$, the probability for value
$x_{i}$ is
\renewcommand{\theequation}{2.14}
\begin{equation}
p\,(x_{i}|\hat{x},\,v)\,
=Z^{-1}\, {\rm Tr} \{\ket{x_{i}}\,\bra{x_{i}}\exp\,(-\beta\,H)\}
= 
\frac{\braket{x_{i}|\E^{-\beta\,H}}{x_{i}}}{Z}
\end{equation}
with partition function
\renewcommand{\theequation}{2.15}
\begin{equation}
Z\,={\rm Tr} \exp\,(-\beta\,H)\,=\,\int dx\ \braket{x|\exp\,(-\beta\,H)}{x}
\end{equation}
where we have dropped the label $\xi$ to simplify notation. For $N$
repeated measurements of
$\hat{x}$ with results $x_{i}$, $i\,=\,1,\,...\,N$, one has under the above
assumptions of independent measurements
\renewcommand{\theequation}{2.16}
\begin{equation}
p\,(x_{T}|O_{T},\,V)\,=\,\prod\limits_{i}\,p\,(x_{i}|\hat{x},\,v)\,=\,
\prod\limits_{i}\,\left[\braket{x_{i}|\E^{-\beta\,H}}{x_{i}} \, Z^{-1}\right]\,
\,.
\end{equation}
Combining eqs. (2.10), (2.11) and (2.16) leads to the posterior
\renewcommand{\theequation}{2.17}
\begin{equation}
p\,(V|D)\,\sim\,\frac{1}{Z^{N}}\,\left(\prod\limits_{i}\,\braket{x_{i}
|\E^{-\beta\,H}}{x_{i}}\right)\,
\exp\,\left(-\frac{\gamma}{2}\,\Gamma\,[v]\right)\,=\,
\exp\,(-E\,(V|D))
\end{equation}
with energy functional
\renewcommand{\theequation}{2.18}
\begin{equation}
E\,(V|D)\,=\,-\sum\limits_{i}\,\ln\,\braket{x_{i}|\E^{-\beta\,H}}{x_{i}}\,
+\,
N\,\ln\,Z\,+\,\frac{\gamma}{2}\,\Gamma\,[v]\,\,,
\end{equation}
functional $\Gamma\,[v]$ defined in eq. (2.11). The corresponding
stationarity equations (2.9) in explicit form
\renewcommand{\theequation}{2.19}
\begin{equation}
-\sum\limits_{i\,=\,1}^{N}\,\frac{\frac{\delta}{\delta\,v\,(x)}\,
\braket{x_{i}|\E^{-\beta\,H}}{x_{i}}}{\braket{x_{i}|\E^{-\beta\,H}}{x_{i}}}
\,+\,
\frac{N}{Z}\,\frac{\delta\,Z}{\delta\,v\,(x)}\,+\,\frac{\gamma}{2}\,\frac
{\delta\,\Gamma}{\delta\,v\,(x)}\,=\,0\,\,,
\label{2.19}
\end{equation}
with
\renewcommand{\theequation}{2.20}
\begin{equation}
\frac{1}{2}\frac{\delta\,\Gamma}{\delta\, v\,(x)} = \,K\,(v\,(x)\,-\,v_{0}
\,(x))\,\,,
\end{equation}
determine the potential $v\,(x)$.

In a series of papers \cite{5}, 
eqs. (2.19) have been studied successfully in energy
representation
for a variety of choices for prior $p\,(V)$. This requires solving the
Schr\"odinger equation
$H\,\ket{\phi_{\alpha}}\,=\,E_{\alpha}\,\ket{\phi_{\alpha}}$, $\braket
{\phi_{\alpha}}{\phi_{\beta}}\,=\,\delta_{\alpha\beta}$, which allows to
calculate the functional derivatives $\delta_{v}\,E_{\alpha}$ and
$\delta_{v}\,\phi_{\alpha}$ 
(cf. appendix) 
needed in eqs. (2.19). In the following
sections we shall apply
the path integral formulation of quantum mechanics in order to study the
semiclassical as well as
classical regimes.

\renewcommand{\thesection}{3}
\section{Likelihood in path integral representation}
The matrix elements appearing in eq. (2.16) can be written as path
integrals \cite{6}
\renewcommand{\theequation}{3.1}
\begin{equation}
\braket{x_{i}|\E^{-\beta\,H}}{x_{i}}\,=\int\limits_{q\,(0)\,=\,x_{i}}^{q\,
(\beta\,\hbar)\,=\,x_{i}}
Dq\,(\tau)\,\exp\,\Biggl\{-\frac{1}{\hbar}\int\limits_{0}^{\beta\,\hbar}
d\tau\left(
\frac{m}{2}\,(dq\,/\,d\tau)^{2}\,+\,v\,(q)\right)\Biggr\}\,\,.
\label{3.1}
\end{equation}
They are related to those of the time development operator of quantum
mechanics by Wick rotation in the complex time plane. 
The corresponding variable
transformation
\renewcommand{\theequation}{3.2}
\begin{equation}
t\,=\,-i\,\tau
\end{equation}
replaces real time $t$ by imaginary time $\tau$ and velocity $dq\,/\,dt
$ by
\renewcommand{\theequation}{3.3}
\begin{equation}
\frac{dq}{d\tau} = \,-i\frac{dq}{dt}\,,
\end{equation}
inducing a change of sign in the kinetic energy term.

Representation (3.1) is understood as abbreviation of an infinite
dimensional integral when dividing the interval $[0,\,\beta\,\hbar]$ into
equidistant segments of length $\varepsilon\,=\,\beta\,\hbar\,/\,M$, coding
the path $q\,(\tau)$
at discrete points $\tau_{k}\,=\,\varepsilon\,k$ by $q_{k}\,=\,q\,
(\tau_{k})$ and taking the limit
$M\,\to\,\infty$
\renewcommand{\theequation}{3.4}
\begin{eqnarray}
\braket{x_{i}|\E^{-\beta\,H}}{x_{i}}
& = & \!\!\!
\begin{array}{c} \lim\vspace*{-2mm}\\ \scriptstyle{M\to\infty}\end{array}
\left(\frac{m}{2\pi\hbar\varepsilon}\right)^{\frac{M}{2}}\int
\Biggl(\prod\limits_{k=1}^{M-1} dq_{k}\Biggr)\nonumber\\
& &
\times\,\exp\Biggl\{-\frac{\varepsilon}{\hbar}\sum\limits_{
k=1}^{M}\!\left(\!\frac{m}{2}\left[\frac{q_{k}-q_{k-1}}{\varepsilon}\right]
^{2}\!\!+v\,(q_{k})\!\right)\!\Biggr\}\nonumber\\
& = &
\int\limits_{q\,(0)\,=\,x_{i}}^{q\,(\beta\,\hbar)\,=\,x_{i}} Dq\,(\tau)\,
\exp\,\left\{
-\frac{1}{\hbar}\,S\,[q]\right\}\,\,,
\end{eqnarray}
with
\renewcommand{\theequation}{3.5}
\begin{equation}
\int\limits_{q\,(0)\,=\,x_{i}}^{q\,(\beta\,\hbar)\,=\,x_{i}} Dq\,(\tau)\,
=
\begin{array}{c} \lim\vspace*{-2mm}\\ \scriptstyle{M\to\infty}\end{array}
\int
\left(\prod\limits_{k\,=\,1}^{M\,-\,1} dq_{k}\right)\,\left(\frac{m}{2\,
\pi\,\hbar\,\varepsilon}\right)
^{\frac{M}{2}}
\end{equation}
and Euclidean action
\renewcommand{\theequation}{3.6}
\begin{equation}
S\,[q]\,=\,\int\limits_{0}^{\beta\,\hbar}\,d\tau\,\left[
\frac{m}{2}\,\dot{q}^{2}\,+\,v\,(q)\right]
\end{equation}
where we have introduced $\dot{q}\,=\,dq/d\tau$ for short. The
boundary
values are fixed, $q_{0}\,=\,q_{M}\,=\,x_{i}$. The partition function $Z$
as trace in coordinate space can
be written as path integral over {\em all} periodic functions $q\,(\tau)$
of fixed period
$\beta\,\hbar$
\renewcommand{\theequation}{3.7}
\begin{eqnarray}
Z = {\rm Tr} (\E^{-\beta\,H})
& = &
\int dx\,\braket{x|\E^{-\beta\,H}}{x}\,=\int dx\int\limits_{q\,(0)\,
=\,x}^{q\,(\beta\,\hbar)\,=\,x}
Dq\,(\tau)\,\exp\left\{-\frac{1}{\hbar}\,S\,[q]\right\}\nonumber\\
& = &
\int\limits_{q\,(0)\,=\,q\,(\beta\,\hbar)}\!\!\!\!\!\!Dq\,(\tau)\,\exp\,
\left\{-\frac{1}{\hbar}\,S\,[q]\right\}\,\,.
\label{like1}
\end{eqnarray}

The path integral for
$\braket{x|\exp\,(-\beta\,H)}{x}$, in lowest order stationary phase
approximation, is given by
\renewcommand{\theequation}{3.8} 
\begin{equation}
\braket{x|\exp\,(-\beta\,H)}{x}\,
= A_x \,\exp\,\left(-\frac{1}{\hbar}
\,S\,[q_{x}]\right)
\label{315}
\end{equation}
with 
$q_{x}\,(\tau)$ being the solution of the classical equations of motion
(\ref{310})
with boundary conditions
\renewcommand{\theequation}{3.9} 
\begin{equation}
q_{x}\,(0)\,=\,q_{x}\,(\beta\,\hbar)\,=\,x\,\,,
\label{316}
\end{equation}
and factor $A_x$ comprising the quadratic fluctuations 
around the classical path $q_x\,(\tau)$
(see section 6, eqs. (\ref{5.7}), (\ref{5.9})).
The $x$-integral in $Z$, eq. (3.7), can also be treated in
stationary phase approximation. The action
$S\,[q]$ depends on $x$ through the boundary values of $q\,(\tau)$, and its
derivative with respect to the upper (lower) boundary
value of $q$ yields the corresponding momentum $\begin{array}{c}
\scriptstyle{+}\vspace*{-2mm}\\ \scriptstyle{(-)}\end{array} p$. The
stationarity condition
for $S$ thus poses the additional boundary condition
\renewcommand{\theequation}{3.10} 
\begin{equation}
p\,(\beta\,\hbar)\,-\,p\,(0)\,=\,0\,\,.
\label{317}
\end{equation}
Hence one has to find $x_{0}$ such that the solutions of the classical
equations
of motion (\ref{310}) fulfill 
boundary conditions for both coordinate $q\,(\tau)$ and
velocity $\dot{q}\,(\tau)$,
\renewcommand{\theequation}{3.11} 
\begin{equation}
q\,(0)\,=\,q\,(\beta\,\hbar)\,=\,x_{0}\quad \text{and}\quad
\dot{q}\,(0)\,=\,\dot{q}\,(\beta\,\hbar)\,\,.
\label{318}
\end{equation}
Then
\renewcommand{\theequation}{3.12} 
\begin{equation}
Z\,
= A_0 \exp\,\left\{-\frac{1}{\hbar}\,S\,[q_{x_{0}}]\right\}
\label{319}
\end{equation}
in lowest order stationary phase approximation,
with $A_0$ the analogue of $A_x$,  eq.\ (\ref{315}).

\renewcommand{\thesection}{4}
\section{Maximum posterior in stationary phase approximation}
In the representation 
(\ref{3.1}), (\ref{like1}), 
the posterior density reads
\renewcommand{\theequation}{4.1}
\begin{equation}
p\,(V|D)\,\sim\,\int\,\left(\prod\limits_{i\,=\,1}^{N}\,Dq_{i}\,(\tau)
\right)\,
\exp\,\left\{-\frac{1}{\hbar}\,\sum\limits_{i\,=\,1}^{N}\,F_i\,[q_{i},\,v]
\right\}
\end{equation}
with total action
\renewcommand{\theequation}{4.2}
\begin{equation}
F_i\,[q_{i},\,v]\,=\,S\,[q_{i},\,v]\,+\,\hbar \,\ln\,Z\,[v]\,+\,
\frac{\hbar\,\gamma}{2N}\,\Gamma\,[v]\,\,,\end{equation}
inserting eqs. (3.4) and (3.6) into (2.17) and (2.18). Note that to each
data point $x_{i}$ is assigned its own path integral.

Following the reasoning in section 2 for the $v$-integration we shall
treat the integrals (3.4) in stationary phase approximation, looking for
paths
$q\,(\tau)$ which minimize the action $S\,[q]$ and account for the main
contribution to the integrals. The corresponding stationarity equations,
\renewcommand{\theequation}{4.3} 
\begin{equation}
0\,=
\frac{\delta\,S}{\delta\,q_{i}} = \,-m\,\ddot{q}_{i}\,+\,
\frac{d}{dq_{i}}\,v\,(q_{i}) \quad \text{for} \quad i\,=\,1,\,...\,,\,N
\label{310}
\end{equation}
with boundary conditions
\renewcommand{\theequation}{4.4} 
\begin{equation}
q_{i}\,(0)\,=\,q_{i}\,(\beta\,\hbar)\,=\,x_{i}
\label{311}
\end{equation}
are the classical equations of motion for a fictitious particle
of mass $m$ in the inverted potential $-v\,(q)$ with boundary conditions
determined by the data points $x_{i}$. Their solutions serve as starting
points for a quantum
mechanical expansion.

For each path $q_{i}\,(\tau)$ the energy
\renewcommand{\theequation}{4.5} 
 \begin{equation}
E_{i}\,=\,\frac{1}{2}\,m\,\dot{q}^{2}_{i}\,-\,v\,(q_{i})
\label{312}
\end{equation}
is conserved. Equations (\ref{310}), (\ref{311}) 
have to be solved simultaneously
with the stationarity equations (\ref{2.19}), explicitly
for $F=\sum_i F_i$:
\renewcommand{\theequation}{4.6} 
\begin{equation}
\begin{array}{c}
\displaystyle{
0\,=\frac{\delta\,F}{\delta\,v\,(x)} 
= \,\sum\limits_{i\,=\,1}^{N}\,
\int\limits_{0}^{\beta\,\hbar}
d\tau\,\delta\,(q_{i}\,(\tau)\,-\,x)\,-\,\beta\,\hbar\,N\,\braket{x|\frac
{\E^{-\beta\,H}}{Z}}{x}}\\
\mbox{}\\
\displaystyle{
+\,\gamma\,\hbar\,\int\,dx^{\prime}\,K\,(x,\,x^{\prime})\,(v\,(x^{\prime})\,-\,v_{0}
\,(x^{\prime}))}
\end{array}
\label{313}
\end{equation}
for the choice (2.10) of the prior $p\,(V)$. 
For the derivative of $\ln\,Z$ we refer to the appendix,
eq.\ (\ref{A.10}).

The integral in the first term 
of (\ref{313}) 
over $\delta$-distributions can be
evaluated, for simple zeroes of the arguments,
with the help of eq. (\ref{312}),
\renewcommand{\theequation}{4.7}
\begin{eqnarray}
\int\limits_{0}^{\beta\,\hbar} d\tau\,\delta\,(q_{i}\,(\tau)\,-\,x)
& = &
\sum\limits_{j_{i}\,=\,1}^{n_{i}\,(x)}\,\frac{1}{|\dot{q}_{i}\,
(\tau_{j_{i}})|}\,=\,
\sum\limits_{j_{i}\,=\,1}^{n_{i}\,(x)}\,\frac{1}{
\sqrt{\frac{2}{m}\,(E_{i}\,+\,v\,(q_{i}\,(\tau_{j_{i}})))}}\nonumber\\
& = &
\frac{n_{i}\,(x)}{\sqrt{\frac{2}{m}\,(E_{i}\,+\,v\,(x))}},
\label{314}
\end{eqnarray}
$n_{i}\,(x)$ being the number of times $\tau_{j_{i}}$ with $q_{i}\,
(\tau_{j_{i}})\,=\,x$,
$0\,\leq\,\tau_{j_{i}}\,\leq\,\beta\,\hbar$. 
In the second term of (\ref{313}),
the path integral for
$\braket{x|\exp\,(-\beta\,H)}{x}$ is in lowest order stationary phase
approximation given by (\ref{315}), and analogously $Z$ by (\ref{319}).

A compact and instructive form of condition (\ref{313}) is obtained by
multiplying with some arbitrary observable $f\,(x)\,/\,N\,\beta\,\hbar$ and
integrating over $x$,
\renewcommand{\theequation}{4.8}
\begin{eqnarray}
0
& = &
\frac{1}{N\,\beta\,\hbar}\sum\limits_{i\,=\,1}^{N}\int\limits_{0}^{\beta\,
\hbar} d\tau\,f\,(q_{i}\,(\tau))\,-\int dx\,f\,(x)\,
\braket{x|\frac{\E^{-\beta\,H}}{Z}}{x}\nonumber\\
& &
+\,\frac{\gamma}{N\,\beta}\int dx\,dx^{\prime}\,f\,(x)\,K(x,\,x^{\prime})
\,[v\,(x^{\prime})\,-\,
v_{0}\,(x^{\prime})]
\\ \nonumber
& = &
\frac{1}{N}\sum\limits_{i\,=\,1}^{N}\,\overline{f}_{i}\,-\,\langle
f\rangle\,+\,
\frac{\gamma}{N\,\beta}\,\int\,dx\,dx^{\prime}\,f\,(x)\,K(x,\,x^{\prime})\,
[v\,(x^{\prime})\,-\,v_{0}\,(x^{\prime})]\,\,,
\label{320}
\end{eqnarray}
where $\overline{f}_{i}$ denotes the mean of $f$ with respect to
(imaginary) time
$\tau$ along path $q_{i}\,(\tau)$ and $\langle f\rangle$ the thermal
expectation value
of observable $f$. Condition (\ref{320}) reminds of the ergodic theorem of
statistical mechanics \cite{7}
concerning time and ensemble average, there are, however,
differences in three respects:
\begin{enumerate}
\item
the time average in (\ref{320}) is over a finite interval only,
\item
paths $q_{i}\,(\tau)$ refer to boundary conditions (\ref{311}) rather than to
initial conditions for
$q\,(\tau)$, $\dot{q}\,(\tau)$, and
\item
the prior gives a contribution to (\ref{320}), non-zero in general, in
contradiction to the ergodic theorem.
\end{enumerate}
In the high temperature limit, $\beta\,\to\,0$, the prior term dominates
condition (\ref{320}),
as expected, since the first two terms in (\ref{320}) become $\beta
$-independent. Prior
knowledge $p\,(V)$ completely determines the maximum posterior solution. In
contrast, the prior on $v$ becomes
negligeable at low temperature, corresponding to large $\beta$-values, and
the first
two terms of (\ref{320}) fulfill the ergodic theorem. In fact, the potential
\renewcommand{\theequation}{4.9}
\begin{equation}
v\,(x)\,=
-\lim_{a\to\infty}
a\,\sum\limits_{i\,=\,1}^{N}\,\delta\,(x\,-\,x_{i})
\label{321}
\end{equation}
is a solution of (\ref{313}), if the prior can be neglected. For the
corresponding classical potential
$-v\,(q)\,=
+\lim_{a\to\infty}
a\,\sum\limits_{i\,=\,1}^{N}\,
\delta\,(q\,-\,x_{i})$ the equations of motion (\ref{310}) with boundary
conditions (\ref{311})
have unstable solutions
\renewcommand{\theequation}{4.10} 
\begin{equation}
q_{i}\,(\tau)\,=\,x_{i}\,\,,
\label{322}
\end{equation}
and the first term in (\ref{313}) reads
\renewcommand{\theequation}{4.11} 
\begin{equation}
\sum\limits_{i\,=\,1}^{N}\int\limits_{0}^{\beta\,\hbar}d\tau\,
\delta\,(q_{i}\,(\tau)-x) =
\int\limits_{0}^{\beta\,\hbar} d\tau \sum\limits_{i\,=\,1}^{N}
\delta\,(x_{i} - x) =
\beta\,\hbar \sum\limits_{i\,=\,1}^{N} \delta\,(x_{i} - x)\,\,.
\label{323}
\end{equation}
For large $\beta$-values
\renewcommand{\theequation}{4.12} 
\begin{equation}
\braket{x|\E^{-\beta\,H}}{x}\quad \to \quad
\sum\limits_{i\,=\,1}^{N}\,\braket{x}{\varphi_{0i}}\,\braket
{\varphi_{0i}}{x}
\E^{-\beta E_0}
\label{324}
\end{equation}
where the $N$-fold degenerate quantum ground state $\braket{x}
{\varphi_{0i}}$
is strongly localized by potential $v\,(x)$, 
eq. (\ref{321}), around the data
points
$x_{i}$ such that
\renewcommand{\theequation}{4.13} 
\begin{equation}
|\braket{x}{\varphi_{0i}}|^{2}\quad \to \quad \delta\,(x\,-\,x_{i})
\label{325}
\end{equation}
in proper normalization. Hence, 
with $Z\,=\,\sum\limits_{i\,=\,1}^{N}\,\E^{-\beta E_0}\,
=\,N\E^{-\beta E_0}$, the second
term in (\ref{313}),
\renewcommand{\theequation}{4.14} 
\begin{equation}
\frac{1}{Z}\,\braket{x|\E^{-\beta\,H}}{x}\quad \to\quad
\frac{1}{N}\,\sum\limits_{i\,=\,1}^{N}\,\delta\,(x\,-\,x_{i})\,\,,
\label{326}
\end{equation}
cancels the first one, eq. (\ref{323}).

So far we have restricted ourselves to position measurements. If given data
refer
to other observables, one can use closure relations to calculate the
required matrix elements in those
observables while retaining the above path integral formalism. A typical
example would be particle momenta rather than positions.
In this case, Fourier transformation leads to
\renewcommand{\theequation}{4.15} 
\begin{eqnarray}
\braket{\tilde{p}|\E^{-\beta\,H}}{\tilde{p}}
& = &
\int dx\,dx^{\prime}\,\braket{\tilde{p}}{x^{\prime}}\,\braket{x^{\prime}
|\E^{-\beta\,H}}{x}\,
\braket{x}{\tilde{p}}\nonumber\\
& \sim &
\int dx\,dx^{\prime}\int\limits_{q\,(0)\,=\,x}^{q\,(\beta\,\hbar)\,=\,x^{\prime}}
Dq\,(\tau)\,\exp\,\left\{-\frac{1}{\hbar}\,\left(S\,[q]\,+\,i\,\tilde{p}
\,(x^{\prime}\,-\,x)\right)\right\}\nonumber\\
& = &
\int Dq\,(\tau)\,\exp\,\left\{-\frac{1}{\hbar}\, \Big(S\,[q]\,+\,i\,\tilde{p}
\,\big(q\,(\beta\,\hbar)\,-\,q\,(0)\big)\Big) \right\}
\label{327}
\end{eqnarray}
where the integration is over {\it all} 
path $q\,(\tau)$ of the interval $[0,\, \beta\,\hbar]$. 
Integral (\ref{327}) may then be calculated 
in saddle point approximation. 
Before presenting a numerical
case study to show that the above formalism is feasible in practice, we
will study an alternative
approach to the path integral representation.

\renewcommand{\thesection}{5}
\section{Maximum posterior stationarity equations in path integral
representation}
In this section we start from the general stationarity equations (2.19) for
the posterior, restricting
ourselves to the case of position measurements for the sake of
definiteness. The terms in (2.19) which
result from the likelihood can all be expressed in terms of the basic path
integral (3.1). As the matrix
elements $\braket{x_{i}|\E^{-\beta\,H}}{x_{i}}$ and partition function $Z\,
=\,
\int dx\,\braket{x|\E^{-\beta\,H}}{x}$ need no further comment, 
we will directly
proceed to evaluate their derivatives with
respect to $v$.

According to (3.7), the functional derivative of $Z$ leads to
\renewcommand{\theequation}{5.1}
\begin{eqnarray}
\frac{\delta\,Z}{\delta\,v\,(x^{\prime})}
& = &
-\frac{1}{\hbar}\,\int\limits_{q\,(0)\,=\,q\,(\beta\,\hbar)}\!\!\!\!Dq\,
(\tau)\,\exp\,\left\{
-\frac{1}{\hbar}\,S\,[q]\right\}\int\limits_{0}^{\beta\,\hbar} d\tau^{\prime}
\,\delta\,
(q\,(\tau^{\prime})\,-\,x^{\prime})\nonumber\\
& = &
-\frac{1}{\hbar}\int\limits_{0}^{\beta\,\hbar} d\tau^{\prime}
\int\limits_{q\,(0)\,=\,q\,(\beta\,\hbar)}\!\!\!\!
Dq\,(\tau)\,\exp\left\{-\frac{1}{\hbar}\,S\,[q]\right\}\delta\,(q\,
(\tau^{\prime})\,-\,x^{\prime})\,,
\nonumber\\
\label{4.1}
\end{eqnarray}
interchanging the order of integration in the second step. To evaluate the
above path
integral we observe that action $S\,[q]$ and measure $Dq\,(\tau)$ are
invariant under cyclic shift
of each path $q\,(\tau)$ around some arbitrary value $\tau^{\prime}$. This is
displayed in Fig.~1 where the
two paths $q_{1}\,(\tau)$, $q_{2}\,(\tau)$ cover the same set of values in
the interval $[0,\,\beta\,\hbar]$ and thus
generate the same value for the integral $\int d\tau\,v\,(q\,(\tau))$.
The same reasoning holds
for the derivatives $\dot{q}_{1}$, $\dot{q}_{2}$, hence $\int d\tau\,
(\dot{q}\,(\tau))^{2}$ also
has the same value for the two paths. Therefore the path integral 
in (\ref{4.1}), running over
all periodic paths which go through point 
$x^{\prime}$ at time $\tau^{\prime}$, can
be expressed as the
path integral over all paths which start and end at $x^{\prime}$:
\renewcommand{\theequation}{5.2}
\begin{eqnarray}
&&\int\limits_{q\,(0)\,=\,q\,(\beta\,\hbar)}\!\!\!\! Dq\,(\tau)\,\exp\,
\left\{-\frac{1}{\hbar}\,S\,[q]\right\}
\delta\,(q\,(\tau^{\prime})\,-\,x^{\prime})
\nonumber\\
&&=
\int\limits_{q\,(0)\,=\,x^{\prime}}^{q\,(\beta\,\hbar)\,=\,x^{\prime}}\!\!\!\!
Dq\,(\tau)\,\exp\,\left\{-\frac{1}{\hbar}\,S\,[q]\right\}\nonumber\\
&&=  
\braket{x^{\prime}|\E^{-\beta\,H}}{x^{\prime}}\,\,,
\label{4.2}
\end{eqnarray}
using eq. (3.1). Note that (\ref{4.2}) holds independent of the choice of
$\tau^{\prime}$. The remaining integral
in (\ref{4.1}) is then trivial, 
$\int\limits_{0}^{\beta\,\hbar}d\tau^{\prime}\,=\,
\beta\,\hbar$, confirming the expected result
\renewcommand{\theequation}{5.3}
\begin{equation}
\frac{\delta\,Z}{\delta\,v\,(x^{\prime})}
= \,\frac{\delta}{\delta\,v\,(x^{\prime})}\int
dx\,
\braket{x|\E^{-\beta\,H}}{x}\,=\,-\beta\,\braket{x^{\prime}
|\E^{-\beta\,H}}{x^{\prime}}\,\,.
\label{4.3}
\end{equation}
\vspace*{1mm}

In the functional derivative of matrix elements $\braket{x_{i}
|\E^{-\beta\,H}}{x_{i}}$ with respect to $v$,
\renewcommand{\theequation}{5.4}
\begin{eqnarray}
\frac{\delta}{\delta\,v\,(x^{\prime})}\,\braket{x_{i}|\E^{-\beta\,H}}{x_{i}}
& = &\!\!\!\!
-\frac{1}{\hbar} \int\limits_{0}^{\beta\,\hbar}d\tau^{\prime}
\!\!\!\!
\int\limits_{q\,(0)\,=\,x_{i}}^{q\,(\beta\,\hbar)\,=\,x_{i}}\!\!
\!\!\!\!
Dq\,(\tau)\,\exp\,\left\{-\frac{1}{\hbar}\,S\,[q]\right\}\,\delta\,(q\,
(\tau^{\prime})\,-\,x^{\prime})
\nonumber\\
\!\!\!\!& = &\!\!\!\!
-\frac{1}{\hbar} \int\limits_{0}^{\beta\,\hbar} d\tau^{\prime}
\!\!\!\!
\int\limits_{\begin{array}{c}
\scriptstyle{q\,(0)\,=\,x_{i}}\\ 
\scriptstyle{q\,(\tau^{\prime})\,=\,x^{\prime}}
\end{array}}^{q\,(\beta\,\hbar)\,=\,x_{i}} 
\!\!\!\!
Dq\,(\tau)\,
\exp\,\left\{-\frac{1}{\hbar}\,S\,[q]\right\}\,\,,
\label{4.4}
\end{eqnarray}
the path integral is split into two separate integrals according to Fig.~2.
Taking into account the
boundary conditions for the paths $q_{1}\,(\tau)$, $q_{2}\,(\tau)$ on the
$\tau$-axis, one obtains, under the $\tau^{\prime}$-integral, a product of
non-diagonal matrix elements
of the statistical operator at different temperatures,
\renewcommand{\theequation}{5.5}
\begin{equation}
\begin{array}{c}
\displaystyle{
\frac{\delta\,\braket{x_{i}|\E^{-\beta\,H}}{x_{i}}}{\delta\,v\,(x^{\prime})} =}
\\
\mbox{}\\
\displaystyle{
-\frac{1}{\hbar}\!\!\!\int\limits_{0}^{\beta\,\hbar} d\tau^{\prime}
\int\limits_{q_{2}\,(\tau^{\prime})\,=\,x^{\prime}}^{q_{2}\,
(\beta\,\hbar)\,=\,x_{i}}\!\!\!Dq_{2}\,(\tau)\,\exp\left\{-\frac{1}
{\hbar} S_{2}\,[q_{2}]\right\}\!\!
\int\limits_{q_{1}\,(0)\,=\,x_{i}}^{q_{1}\,(\tau^{\prime})\,=\,x^{\prime}}\!\!
\!Dq_{1}\,(\tau)\,\exp \left\{
-\frac{1}{\hbar}\,S_{1}\,[q_{1}]\right\}}\\
\mbox{}\\
\displaystyle{
=\,-\frac{1}{\hbar} \int\limits_{0}^{\beta\,\hbar} d\tau^{\prime}\,
\braket{x_{i}|\exp\,\left\{
-\left(\beta\,-\,\frac{\tau^{\prime}}{\hbar}\right)\,H\right\}}{x^{\prime}}\,
\braket{x^{\prime}|\exp\,\left\{-\frac{\tau^{\prime}}{\hbar}\,H\right\}}{x_{i}}}\\
\mbox{}\\
\displaystyle{
=\,-\int\limits_{0}^{\beta}\,d\beta^{\prime}\,\braket{x_{i}|\exp\,\{-(\beta\,
-\,\beta^{\prime})\,H\}}{x^{\prime}}\,
\braket{x^{\prime}|\exp\,\{-\beta^{\prime}\,H\}}{x_{i}} }
\end{array}
\label{4.5}
\end{equation}
where the full action $S\,[q]$ is split into parts,
\renewcommand{\theequation}{5.6}
\begin{equation}
\begin{array}{c}
\displaystyle{
S_{1}\,[q_{1}]\,= \int\limits_{0}^{\tau^{\prime}} d\tau\,\left(
\frac{m}{2}\,\dot{q}_{1}^{2}\,+\,v_{1}\,(q_{1})\right)}\\
\mbox{}\\
\text{and}\\
\mbox{}\\
\displaystyle{
S_{2}\,[q_{2}]\,=\int\limits_{\tau^{\prime}}^{\beta\,\hbar} d\tau\,\left(
\frac{m}{2}\,\dot{q}^{2}_{2}\,+\,v_{2}\,(q_{2})\right)\,\,.}
\end{array}
\label{4.6}
\end{equation}
The results (\ref{4.3}), (\ref{4.5}) and (\ref{4.6}) 
obtained from the basic path integral
(3.1) are still exact, in
particular they are strictly equivalent to the respective expressions in
energy representation (see appendix). With
the above formulae, stationarity equations (2.19) read
\renewcommand{\theequation}{5.7}
\begin{equation}
\begin{array}{c}
\displaystyle{
\sum\limits_{i\,=\,1}^{N} \int\limits_{0}^{\beta}\,d\beta^{\prime}\,\frac{
\braket{x_{i}|\exp\,(-(\beta\,-\,\beta^{\prime})\,H)}{x^{\prime}}\,
\braket{x^{\prime}|\exp\,(-\beta^{\prime}\,H)}{x_{i}}}{\braket{x_{i}|\exp\,
(-\beta\,H)}{x_{i}}}\,-\,
N\,\beta\,\braket{x^{\prime}|\frac{\E^{-\beta\,H}}{Z}}{x^{\prime}}}\\
\mbox{}\\
\displaystyle{
=\,-\frac{\gamma}{2}
\frac{\delta\,\Gamma}{\delta\,v\,(x^{\prime})} =\,
-\gamma\,K\,(v\,(x^{\prime})\,-\,
v_{0}\,(x^{\prime}))}
\label{4.7}
\end{array}
\end{equation}
for the prior of eqs. (2.10), (2.11).

In the following sections we shall study approximation schemes for the
above derivatives 
(\ref{4.3}) and (\ref{4.5}).
It will be shown under what assumptions the exact (quantum mecha\-nical)
result (\ref{4.7}) 
approaches the approximate (semiclassical) form (\ref{313}), 
and how to
find corrections to (\ref{313}) taking into account quantum fluctuations around
the classical paths $q_{i}\,(\tau)$ 
of equations (\ref{310}), (\ref{311}).

\renewcommand{\thesection}{6}
\section{Quadratic fluctuations}
Matrix elements
\renewcommand{\theequation}{6.1}
 \begin{equation}
\braket{x|\E^{-\beta\,H}}{x}\,= \int\limits_{q\,(0)\,=\,x}^{q\,(\beta\,
\hbar)\,=\,x}
Dq\,(\tau)\,\exp\,\left\{-\frac{1}{\hbar}\,S\,[q]\right\}
\label{5.1}
\end{equation}
can be handled by standard techniques, starting from the stationarity
equations (\ref{310}), (\ref{311}). 
They read, dropping the label ${i}$ and using $v^{\prime}\,=\,
dv/\,dq_{x}$ for simplicity of notation,
\renewcommand{\theequation}{6.2}
\begin{equation}
0\,=\frac{\delta\,S}{\delta\,q} =\,-m\,\ddot{q}\,+\,
v^{\prime}\,(q) \quad \text{with} \quad q\,(0)\,=\,q\,(\beta\,\hbar)\,=\,x\,\,.
\label{5.2}
\end{equation}
The stationary solutions $q_{x}\,(\tau)$ of (\ref{5.2}) yield the main
contribution to the path integral (\ref{5.1});
fluctuations around these solutions $q_{x}\,(\tau)$ can be taken care of by
a variable transformation
\renewcommand{\theequation}{6.3}
\begin{equation}
q\,(\tau)\,=\,q_{x}\,(\tau)\,+\,r\,(\tau) \quad \text{with} \quad
r\,(0)\,=\,r\,(\beta\,\hbar)\,=\,0\,\,.
\label{5.3}
\end{equation}
Assuming that only small deviations of $q\,(\tau)$ from $q_{x}\,(\tau)$ are
important for the integral (\ref{5.1}), we approximate
\renewcommand{\theequation}{6.4} 
\begin{equation}
v\,(q)\,=\,v\,(q_{x})\,+\,(q\,-\,q_{x})\,v^{\prime}\,(q_{x})\,+\,
\frac{1}{2}\,(q\,-\,q_{x})^{2}\,v^{''}\,(q_{x})
\label{5.4}
\end{equation}
and find for the action
\renewcommand{\theequation}{6.5} 
\begin{eqnarray}
S\,[q]
& = &
\int\limits_{0}^{\beta\,\hbar}\,d\tau\,\left(\frac{m}{2}\,\dot{q}^{2}_{x}
\,+\,
\frac{m}{2}\,\dot{r}^{2}\,+\,m\,\dot{q}_{x}\,\dot{r}\,+\,v\,(q_{x}\,+\,r)
\right)\nonumber\\
& = &
\int\limits_{0}^{\beta\,\hbar} d\tau\,\left(\frac{m}{2}\,\dot{q}^{2}_{x}
\,+\,v\,(q_{x})\right)\,+
\int\limits_{0}^{\beta\,\hbar} d\tau\,\left(\frac{m}{2}\,\dot{r}^{2}\,+\,
\frac{1}{2}\,
v^{''}\,(q_{x})\,r^{2}\right)\nonumber\\
& &
+\,\int\limits_{0}^{\beta\,\hbar}\,d\tau\,\Big(m\,\dot{q}_{x}\,\dot{r}\,
+\,v^{\prime}\,(q_{x})\,r\Big)
\label{5.5}
\end{eqnarray}
where the last term vanishes by virtue of (\ref{5.2}), (\ref{5.3}) and partial
integration
\renewcommand{\theequation}{6.6} 
\begin{equation}
\int\limits_{0}^{\beta\,\hbar} d\tau\,[m\,\dot{q}_{x}\,\dot{r}\,+\,
v^{\prime}\,(q_{x})\,r]\,=\,\left[m\,\dot{q}_{x}\,r\right]^{\beta\,\hbar}_{0}\,
=\,0\,\,.
\label{5.6}
\end{equation}
For the additive action (\ref{5.5}), (\ref{5.6}) 
the matrix element (\ref{5.1}) factorizes
\renewcommand{\theequation}{6.7} 
\begin{equation}
\begin{array}{c}
\displaystyle{
\braket{x|\E^{-\beta\,H}}{x}
= A_x \exp\left\{ -\frac{1}{\hbar} \,S\,[q_x]\right\} 
}
\\
\displaystyle{
=\exp\,\left\{-\frac{1}{\hbar}\,S\,[q_{x}]
\right\}
\int\limits_{r\,(0)\,=\,0}^{r\,(\beta\,\hbar)\,=\,0}Dr\,(\tau)}\\
\displaystyle{
\times\,\,
\exp\,\left\{-\frac{1}{\hbar}\int\limits_{0}^{\beta\,\hbar}\!d\tau\,
\left(\frac{m}{2}\,\dot{r}^{2}+\frac{1}{2}\,v^{''}\,(q_{x})\,r^{2}\right)
\right\}}\\
\mbox{}\\
\displaystyle{
=\exp\,\left\{-\frac{1}{\hbar}\,S\,[q_{x}]\right\} 
\int\limits_{r\,(0)\,=\,0}^{r\,(\beta\,\hbar)\,=\,0}Dr\,(\tau)}\\
\displaystyle{
\times\,\,
\exp\,\left\{-\frac{1}{2\,\hbar}\int\limits_{0}^{\beta\,\hbar}\!\!
d\tau\,d\tau^{\prime}\,
r\,(\tau^{\prime})\left(\frac{\delta^{2}\,S\,[q]}{\delta\,q\,(\tau^{\prime})\,
\delta\,q\,(\tau)}\Bigg|_{q\,=\,q_{x}}\right)\,r\,(\tau)\right\} }
\end{array}
\label{5.7}
\end{equation}
with Hesse-matrix
\renewcommand{\theequation}{6.8} 
\begin{equation}
\frac{\delta^{2}\,S\,[q]}{\delta\,q\,(\tau^{\prime})\,\delta\,q\,(\tau)}
\Biggr|_{q\,=\,q_{x}}\,=\,
\delta\,(\tau\,-\,\tau^{\prime})\,\left(-m\,\frac{\partial^{2}}{\partial\,
\tau^{2}}\,+\,v^{''}\,(q_{x}\,(\tau))\right)\,\,,
\label{5.8}
\end{equation}
in approximation (\ref{5.4}). 
For the integral in (\ref{5.7}) to be well-defined, the
Hesse-matrix (\ref{5.8}) has to
be positive-definite. Under the expansion (\ref{5.4}) 
this holds if $v^{''}\,(x) \,>\,0$ for all $x$.

The remaining path integral (\ref{5.7}) can be evaluated by the `shifting
method'. We shall simply recall
the result, known in the literature as van Vleck--formula \cite{8}:
\renewcommand{\theequation}{6.9} 
\begin{eqnarray}
\braket{x|\E^{-\beta\,H}}{x}
&=& A_x \exp\left\{ -\frac{1}{\hbar}\, S\,[q_x]\right\} \nonumber\\
& = &
\left(\frac{2\,\pi\,\hbar}{m}\,\kappa_{x}\,(\beta\,\hbar)\,\kappa_{x}\,(0)
\int\limits_{0}^{\beta\,\hbar} \frac{d\tau}{\kappa^{2}_{x}\,(\tau)}
\right)^{-\frac{1}{2}}
\exp\,\left\{-\frac{1}{\hbar}\,S\,[q_{x}]\right\}\nonumber\\
& = &
\left(-\frac{1}{2\,\pi\,\hbar}\,\frac{\partial^{2}\,S\,[q_{x}]}
{\partial\,x^{2}}\right)^{\frac{1}{2}}\,
\exp\,\left\{-\frac{1}{\hbar}\,S\,[q_{x}]\right\}
\label{5.9}
\end{eqnarray}
with $\kappa_{x}\,(\tau)$ a solution of
\renewcommand{\theequation}{6.10} 
\begin{equation}
v^{''}\,(q_{x}\,(\tau))\,=\,m\,\frac{\ddot{\kappa}_{x}\,(\tau)}{\kappa_{x}
\,(\tau)}\,\,.
\label{5.10}
\end{equation}
If $\dot{q}_{x}$ does not vanish on the path $q_{x}\,(\tau)$, one can
choose
\renewcommand{\theequation}{6.11} 
\begin{equation}
\kappa_{x}\,(\tau)\,=\,\dot{q}_{x}\,(\tau)\,\,,
\label{5.11}
\end{equation}
as is easily seen by differentiating (\ref{5.2}) with respect to $\tau$.
Otherwise we look for a linear
combination of the two linearly independent solutions $\kappa^{(1)}_{x}\,
(\tau)\,=\,
\dot{q}_{x}\,(\tau)$ and
\renewcommand{\theequation}{6.12} 
\begin{equation}
\kappa^{(2)}_{x}\,(\tau)\,=\,\dot{q}_{x}\,(\tau) \int\limits^{\tau}
\frac{d\tau^{\prime}}{(\dot{q}_{x}\,(\tau^{\prime}))^{2}}\,\,.
\label{5.12}
\end{equation}
The latter solution follows from the fact that the Wronski determinant 
of eq. (\ref{5.2}) is constant.


For the partition function $Z$ we use the result (\ref{5.7}), (\ref{5.9}) 
for the matrix element of the statistical operator
\renewcommand{\theequation}{6.13} 
\begin{equation}
Z= 
\int dx^\prime \, A_{x^\prime}
\exp\,\left\{-\frac{1}{\hbar}\,S\,[q_{x^\prime}]\right\},
\label{5.13}
\end{equation}
and the $x^\prime$--integration is done numerically.
Combining (\ref{5.7}) and (\ref{5.13}) 
results in the normalized matrix element of the statistical operator
in coordinate representation
\renewcommand{\theequation}{6.14} 
\begin{equation}
\braket{x|\E^{-\beta\,H}}{x}\,/\,Z\,
=\,
\frac{A_{x} \exp\,\left\{-\frac{1}{\hbar}\,S\,[q_{x}]\right\}}
{\int dx^\prime \, A_{x^\prime}
\exp\,\left\{-\frac{1}{\hbar}\,S\,[q_{x^\prime}]\right\}}.
\label{5.14}
\end{equation}

For large masses $m$, formula (\ref{5.7}) reproduces the result of classical
statistical mechanics.
In this case, the equations of motion (\ref{5.2}) simplify,
\renewcommand{\theequation}{6.15} 
\begin{equation}
\ddot{q}_{x}\,=\,\frac{1}{m}\,v^{\prime}\,(q_{x})\,\to\,0 \quad \text{for} \quad
m\,\to\,\infty\,\,,
\label{5.15}
\end{equation}
and are solved by the static paths
\renewcommand{\theequation}{6.16} 
\begin{equation}
q_{x}\,(\tau)\,=\,x
\label{5.16}
\end{equation}
for the boundary conditions of (\ref{5.2}). Then from (\ref{5.7})
\renewcommand{\theequation}{6.17} 
\begin{equation}
\braket{x|\E^{-\beta\,H}}{x}\to\exp\,(-\beta\,v\,(x))\!\!\!
\int\limits_{r\,(0)\,=\,0}^{r\,(\beta\,\hbar)\,=\,0}\!\!\!
Dr\,(\tau)\,\exp\left\{-\frac{1}{\hbar}\int\limits^{\beta\,\hbar}_{0}\!
\!d\tau\!\left(\frac{m}{2}\dot{r}^{2} +
\frac{1}{2} v^{''}\,(x)\,r^{2}\!\right)\!\right\}
\label{5.17}
\end{equation}
In the limit of large masses, $v^{''}\,(x)\,/\,m\,\to\,0$, the remaining
integral in (\ref{5.17}) becomes independent
of $x$ so that the classical result is obtained:
\renewcommand{\theequation}{6.18} 
\begin{equation}
\braket{x|\E^{-\beta\,H}}{x}\,/\,Z\,=\,\frac{\exp\,(-\beta\,v\,(x))}{\int
dx^{\prime}\,\exp\,(-\beta\,v\,(x^{\prime}))}\,\,.
\label{5.18}
\end{equation}

\renewcommand{\thesection}{7}
\section{Matrix elements of the derivative of the statistical operator with
respect to the potential}
Three variants of approximations for the derivative
\renewcommand{\theequation}{7.1} 
\begin{equation}
\frac{\delta}{\delta\,v\,(x^{\prime})}\,
\braket{x_{i}|\E^{-\beta\,H}}{x_{i}}\,=\,-\frac{1}{\hbar}\!\int\limits_{0}
^{\beta\,\hbar}d\tau^{\prime}\int\limits_{q\,(0)\,=\,x_{i}}^{q\,(\beta\,\hbar)
\,=\,x_{i}}\!\!\!
Dq\,(\tau)\,\exp\,\left\{\!-\frac{1}{\hbar}\,S\,[q]\right\}\,\delta\,(q\,
(\tau^{\prime})\,-\,x^{\prime})
\label{6.1}
\end{equation}
are presented. The general strategy is to find approximations such that in
the logarithmic derivative of the statistical
operator, needed in (2.19), the statistical operator drops out.
\vspace*{1mm}

In the {\bf first approach}, 
we observe that the main contribution to the path
integral
stems from the stationary path $q_{x_{i}}\,(\tau)$, solution of eqs.
(\ref{310}), (\ref{311}).
Hence the distribution 
$\delta\,(q\,(\tau^{\prime})\,-\,x^{\prime})$ {\em under} the
path
integral in (\ref{6.1}) 
may be replaced by $\delta\,(q_{x_{i}}\,(\tau^{\prime})\,
-\,x^{\prime})$, referring to the
stationary path, {\em in front} of the path integral. In this
approximation,
\renewcommand{\theequation}{7.2} 
\begin{equation}
\frac{\delta}{\delta\,v\,(x^{\prime})}\braket{x_{i}|\E^{-\beta\,H}}{x_{i}}
=-\frac{1}{\hbar}\left\{\!
\int\limits_{0}^{\beta\,\hbar}\!d\tau^{\prime}\delta(q_{x_{i}}(\tau^{\prime}) -
x^{\prime})\!\right\}\!\!
\int\limits_{q\,(0)\,=\,x_{i}}^{q\,(\beta\,\hbar)\,=\,x_{i}}
\!\!\!\!\!\!\!\!\!
Dq\,(\tau)\exp\left\{-\frac{1}{\hbar}S\,[q]\right\},
\label{6.2}
\end{equation}
the first term in the stationarity equation (2.19) takes the form
\renewcommand{\theequation}{7.3} 
\begin{equation}
\frac{\frac{\delta}{\delta\,v\,(x^{\prime})}\,\braket{x_{i}
|\E^{-\beta\,H}}{x_{i}}}{
\braket{x_{i}|\E^{-\beta\,H}}{x_{i}}}\,=\,-\frac{1}{\hbar} \int\limits_{0}
^{\beta\,\hbar}
d\tau^{\prime}\,\delta\,(q_{x_{i}}\,(\tau^{\prime})\,-\,x^{\prime})
\label{6.3}
\end{equation}
in agreement with the first term in (\ref{313}). 
One may improve on the result
(\ref{6.3}) by applying approximation (\ref{6.2}) after the
classical action $S\,[q_{x_{i}}]$ has been factored out with the help of
the variable transformation (\ref{5.3}). Then, using (\ref{5.7}),
\renewcommand{\theequation}{7.4} 
\begin{equation}
\begin{array}{c}
\displaystyle{
\frac{\delta}{\delta\,v\,(x^{\prime})}\,\braket{x_{i}|\E^{-\beta\,H}}{x_{i}}\,
=\,
-\frac{\exp\,\left\{-\frac{1}{\hbar}\,S\,[q_{x_{i}}]\right\} }{\hbar}
\int\limits^{\beta\,\hbar}_{0}d\tau^{\prime}
\int\limits_{r\,(0)\,=\,0}^{r\,(\beta\,\hbar)\,=\,0}Dr\,(\tau)}\\
\mbox{}\\
\displaystyle{\times\,\,
\delta\,(q_{x_{i}}\,(\tau^{\prime})\,+\,r\,(\tau^{\prime})\,-\,x^{\prime})\,\exp\,\left\{
-\frac{1}{\hbar}\,S\,[r]\right\}}\\
\mbox{}\\
\displaystyle{
\approx-\frac{\exp\left\{-\frac{1}{\hbar}\,S\,[q_{x_{i}}]\right\}}{\hbar}\!
\int\limits_{0}^{\beta\,\hbar}\!\!
d\tau^{\prime}\,\delta\,(q_{x_{i}}\,(\tau^{\prime})+r_{x_{i}}\,(\tau^{\prime})-x^{\prime})\!
\!\!\!\!\!
\int\limits_{r\,(0)\,=\,0}^{r\,(\beta\,\hbar)\,=\,0}\!\!\!\!\! Dr\,(\tau)
\exp\left\{
-\frac{1}{\hbar}S\,[r]\!\right\} }
\end{array}
\label{6.4}
\end{equation}
where $r_{x_{i}}\,(\tau)$ is the solution of
\renewcommand{\theequation}{7.5} 
\begin{equation}
m\,\ddot{r}_{x_{i}}\,=\,v^{''}\,(q_{x_{i}}\,(\tau))\,r_{x_{i}}\quad
\text{for}\quad
r_{x_{i}}\,(0)\,=\,0\,=\,r_{x_{i}}\,(\beta\,\hbar)\,\,,
\label{6.5}
\end{equation}
and
\renewcommand{\theequation}{7.6} 
\begin{equation}
S\,[r]\,=\int\limits_{0}^{\beta\,\hbar} d\tau\,\left(\frac{1}{2}\,m\,
\dot{r}^{2}\,+\,
\frac{1}{2}\,\,v^{''}\,(q_{x_{i}}\,(\tau))\,r^{2}\right)\,\,.
\label{6.6}
\end{equation}
With (\ref{5.7}),
\renewcommand{\theequation}{7.7} 
\begin{equation}
\braket{x_{i}|\E^{-\beta\,H}}{x_{i}}\,=\,\exp\,\left\{-\frac{1}{\hbar}
\,S\,[q_{x_{i}}]\right\}
\int\limits_{r\,(0)\,=\,0}^{r\,(\beta\,\hbar)\,=\,0}\!\!\!Dr\,(\tau)\,
\exp\,\left\{-\frac{1}{\hbar}\,S\,[r]\right\}\,\,,
\label{6.7}
\end{equation}
one finally has
\renewcommand{\theequation}{7.8} 
\begin{equation}
\frac{\frac{\delta}{\delta\,v\,(x^{\prime})}
\braket{x_{i}|\E^{-\beta\,H}}{x_{i}}
}{
\braket{x_{i}|\E^{-\beta\,H}}{x_{i}}}\,=\,-\frac{1}{\hbar} \int\limits_{0}
^{\beta\,\hbar}
d\tau^{\prime}\,
\delta\,\Big(q_{x_{i}}\,(\tau^{\prime})\,+\,r_{x_{i}}\,(\tau^{\prime})\,
-\,x^{\prime}\Big)\,\,.
\label{6.8}
\end{equation}
\vspace*{1mm}

A {\bf second possibility} 
to evaluate the derivative of the statistical operator
consists of
inserting the spectral representation of the $\delta$-distribution into
(\ref{6.1}). With
$q\,(\tau)\,=\,q_{x_{i}}\,(\tau)\,+\,r\,(\tau)$ one obtains,
in approximation (\ref{5.4}),
for
\renewcommand{\theequation}{7.9} 
\begin{equation}
\begin{array}{c}
\displaystyle{
\frac{\delta}{\delta\,v\,(x^{\prime})}\,\braket{x_{i}|\E^{-\beta\,H}}{x_{i}}
= \,-\frac{1}{\hbar}\,
\exp\,\left\{-\frac{1}{\hbar}\,S\,[q_{x_{i}}]\right\}\!\!\int\limits_{0}
^{\beta\,\hbar}\!\!d\tau^{\prime}\!\!
\!\int\!\frac{d\lambda}{2\,\pi}\exp\left\{i\,\lambda\,(q_{x_{i}}\,
(\tau^{\prime})-x^{\prime})\right\}}\\
\mbox{}\\
\displaystyle{
\times \int\limits_{r\,(0)\,=\,0}^{r\,(\beta\,\hbar)\,=\,0}\!\!\!Dr\,
(\tau)\exp\left\{
-\frac{1}{\hbar}\!\!\int\limits_{0}^{\beta\,\hbar}\!\!d\tau\left
(\frac{m}{2}\,\dot{r}^{2}\,+\,\frac{1}{2}\,v^{''}\,(q_{x_{i}})\,r^{2}-
i\,\lambda\,\hbar\,\delta(\tau-\tau^{\prime})\,r(\tau)\right)\right\}}
\end{array}
\label{6.9}
\end{equation}
a path integral where a $\delta$-like potential appears in the action in
addition to the harmonic potential with
time dependent frequency. Looking for saddle points of the path integral
leads to the inhomogeneous equation of motion
\renewcommand{\theequation}{7.10} 
\begin{equation}
-m\,\ddot{r}_{x_{i}\,\tau^{\prime}\,\lambda}\,(\tau)\,
+\,v^{\prime\prime}(q_{x_{i}}(\tau))
\,r_{x_{i}\,\tau^{\prime}\,\lambda}\,=\,
i\,\lambda\,\hbar\,\delta\,(\tau\,-\,\tau^{\prime})
\label{6.10}
\end{equation}
with boundary conditions
\renewcommand{\theequation}{7.11} 
\begin{equation}
r_{x_{i}\,\tau^{\prime}\,\lambda}\,(0)\,=\,r_{x_{i}\,\tau^{\prime}\,\lambda}\,
(\beta\,\hbar)\,=\,0\,\,.
\label{6.11}
\end{equation}
Note that the solutions of (\ref{6.10}), (\ref{6.11}) 
are both $\lambda$- and $\tau^{\prime}$-dependent. 
The path $r_{x_{i}\,\tau^{\prime}\,\lambda}$ is
intimately related to the Green function $R_{x_{i}}\,(\tau,\,\tau^{\prime})$ of
the operator
$-m\,\frac{d^{2}}{d\tau^{2}}\,+\,v^{''}\,(q_{x_{i}}\,(\tau))$ with the
above boundary conditions,
\renewcommand{\theequation}{7.12} 
\begin{equation}
-m\,\ddot{R}_{x_{i}}\,(\tau,\,\tau^{\prime})\,+\,v^{''}\,(q_{x_{i}}\,(\tau))
\,R_{x_{i}}\,(\tau,\,\tau^{\prime})\,=\,
\delta\,(\tau\,-\,\tau^{\prime})
\label{6.12}
\end{equation}
with
\renewcommand{\theequation}{7.12'} 
\begin{equation}
R_{x_{i}}\,(0,\,\tau^{\prime})\,=\,R_{x_{i}}\,(\beta\,\hbar,\,\tau^{\prime})\,=\,0\,
\,,
\label{6.12'}
\end{equation}
namely
\renewcommand{\theequation}{7.13} 
\begin{equation}
r_{x_{i}\,\tau^{\prime}\,\lambda}\,(\tau)\,=\,i\,\hbar\,\lambda\,R_{x_{i}}\,
(\tau,\,\tau^{\prime})\,\,.
\label{6.13}
\end{equation}
Furthermore, multiplying (\ref{6.10}) 
by $r_{x_{i}\,\tau^{\prime}\,\lambda}\,(\tau)$
and integrating over
$\tau$, one obtains for the quadratic part of the stationary action of
(\ref{6.9})
\renewcommand{\theequation}{7.14} 
\begin{eqnarray}
-\frac{1}{\hbar}\int\limits_{0}^{\beta\,\hbar}\!\!d\tau\left(\frac{1}{2}
\,m\,\dot{r}^{2}_{x_{i}\,\tau^{\prime}\,\lambda}+
\frac{1}{2}\,v^{''}\,(q_{x_{i}}\,(\tau))\,r_{x_{i}\,\tau^{\prime}\lambda}^{2}
\right)\!
& = & -\frac{i\,\lambda}{2}\,r_{x_{i}\tau^{\prime}\lambda}(\tau^{\prime})\nonumber\\
& = &
\frac{\lambda^{2}\hbar}{2}\,R_{x_{i}}\,(\tau^{\prime},\,\tau^{\prime})
\label{6.14}
\end{eqnarray}
Here use has been made of a partial integration, $-\int d\tau\,\dot{r}
^{2}\,=\int d\tau\,r\,\ddot{r}$, 
and of eqs. (\ref{6.10}) and (\ref{6.11}). After
further variable transformation,
\renewcommand{\theequation}{7.15} 
\begin{equation}
r\,(\tau)\,=\,r_{x_{i}\,\tau^{\prime}\,\lambda}\,(\tau)\,+\,l\,(\tau)\,\,,
\label{6.15}
\end{equation}
one finds that in the action of eq. (\ref{6.9})
\renewcommand{\theequation}{7.16} 
\begin{equation}
\begin{array}{c}
\displaystyle{
 \int\limits_{0}^{\beta\,\hbar}\!\!d\tau\,\Bigl(m\,\dot{r}_{x_{i}\,
 \tau^{\prime}\,\lambda}\,(\tau)\,\dot{l}\,(\tau)\,+\,
v^{''}\,(q_{x_{i}}\,(\tau))\,r_{x_{i}\,\tau^{\prime}\,\lambda}\,(\tau)\,l\,
(\tau)\,-\,i\,\hbar\lambda \,l\,(\tau)\,\delta\,(\tau-\tau^{\prime})\Bigr)}\\
\mbox{}\\
=\,
[m\,\dot{r}_{x_{i}\,\tau^{\prime}\,\lambda}\,(\tau)\,l\,(\tau)]^{\beta\,\hbar}
_{0}\,=\,0
\end{array}
\label{6.16}
\end{equation}
so that (\ref{6.9}) reads, under approximation (\ref{6.2}),
\renewcommand{\theequation}{7.17} 
\begin{equation}
\begin{array}{c}
\displaystyle{
\frac{\delta\,\braket{x_{i}|\E^{-\beta\,H}}{x_{i}}}{\delta\,v\,(x^{\prime})}
=-\frac{1}{\hbar}\,
\exp\,\left\{-\frac{1}{\hbar}\,S\,[q_{x_{i}}]\right\}
\!\!\int\limits_{0}^{\beta\,\hbar}
\!\!\!d\tau^{\prime}\,
\Biggl[\int \frac{d\lambda}{2\,\pi}\exp\Biggl\{i\,\lambda\,(q_{x_{i}}\,
(\tau^{\prime})\,+\,}\\
\mbox{}\\
\displaystyle{
r_{x_{i}\,\tau^{\prime}\,\lambda}\,(\tau^{\prime})-x^{\prime})\Biggr\}
\exp\left\{-\frac{1}{\hbar}\,S\,[r_{x_{i}\,\tau^{\prime}\,\lambda}]\right\}
\Biggr]\!\!
\int\limits_{l\,(0)\,=\,0}^{l\,(\beta\,\hbar)\,=\,0}
Dl\,(\tau)\,\exp\,\left\{-\frac{1}{\hbar}\,S\,[l]\right\}\,,}
\end{array}
\label{6.17}
\end{equation}
with abbreviations $S\,[r_{x_{i}\,\tau^{\prime}\,\lambda}]$ 
and $S\,[l]$ defined according to (\ref{6.6}). 
Using (\ref{6.14}), 
the integral over $\lambda$ is of Gaussian type and can be carried
out:
\renewcommand{\theequation}{7.18} 
\begin{equation}
\begin{array}{c}
\displaystyle{
\int \frac{d\lambda}{2\,\pi}\,\exp\,\left\{i\,\lambda\,(q_{x_{i}}\,
(\tau^{\prime})\,+\,
\frac{1}{2}\,r_{x_{i}\,\tau^{\prime}\,\lambda}\,(\tau^{\prime})\,-\,x^{\prime})\right\}}\\
\mbox{}\\
\displaystyle{
= \int\frac{d\lambda}{2\,\pi}\,\exp\,\left\{i\,\lambda\,\left(q_{x_{i}}\,
(\tau^{\prime})\,-\,x^{\prime}\right)\,-\,
\frac{\lambda^{2}\,\hbar}{2}\,R_{x_{i}}\,(\tau^{\prime},\,\tau^{\prime})\right\}}\\
\mbox{}\\
\displaystyle{
=\,\frac{1}{\sqrt{2\,\pi\,\hbar\,R_{x_{i}}\,(\tau^{\prime},\,\tau^{\prime})}}\,\exp\,
\left\{-\frac{(q_{x_{i}}\,(\tau^{\prime})\,-\,x^{\prime})^{2}}{2\,\hbar\,R_{x_{i}}\,
(\tau^{\prime},\,\tau^{\prime})}\right\}\,\,.}
\end{array}
\label{6.18}
\end{equation}
The final result for the logarithmic derivative of $\braket{x_{i}
|\E^{-\beta\,H}}{x_{i}}$ is
\renewcommand{\theequation}{7.19} 
\begin{equation}
\frac{\frac{\delta}{\delta\,v\,(x^{\prime})}\,\braket{x_{i}
|\E^{-\beta\,H}}{x_{i}}}{\braket{x_{i}|\E^{-\beta\,H}}{x_{i}}}\,=\,-
\frac{1}{\hbar}\int\limits_{0}^{\beta\,\hbar}d\tau^{\prime}\,\frac{\exp\,
\left\{-\frac{
(q_{x_{i}}\,(\tau^{\prime})\,-\,x^{\prime})^{2}}{2\,\hbar\,R_{x_{i}}\,(\tau^{\prime},\,
\tau^{\prime})}\right\}}{
\sqrt{2\,\pi\,\hbar\,R_{x_{i}}\,(\tau^{\prime},\,\tau^{\prime})}}\,\,,
\label{6.19}
\end{equation}
inserting
\renewcommand{\theequation}{7.20} 
\[
\braket{x_{i}|\E^{-\beta\,H}}{x_{i}}
=\,
\]
\begin{equation}
\exp\,\left\{-\frac{1}{\hbar}\,S\,[q_{x_{i}}]\right\} 
\exp\left\{-\frac{1}{\hbar}\,S\,[r_{x_{i}\,\tau^{\prime}\,\lambda}]\right\}
\int\limits_{l\,(0)}
^{l\,(\beta\,\hbar)\,=\,0}
Dl\,(\tau)\,\exp\,\left\{-\frac{1}{\hbar}\,S\,[l]\right\}\,
\label{6.20}
\end{equation}
according to (\ref{5.7}). 
In comparison to (\ref{6.3}) and (\ref{6.8}) of the first approach,
the $\delta$-distributions in (\ref{6.3}) and (\ref{6.8})
are replaced in (\ref{6.19}) 
by Gaussians which are normalized with respect to
$x^{\prime}$ and whose
widths are given by the Green functions $R_{x_{i}}\,(\tau,\,\tau)$. Note
that
\renewcommand{\theequation}{7.21} 
\begin{equation}
R_{x_{i}}\,(\tau^{\prime},\,\tau^{\prime})\,
=\int\limits_{0}^{\beta\,\hbar} d\tau\,
\left(\!
\frac{1}{2}\,m\!\left(\frac{\partial\,R_{x_{i}}\,(\tau,\,\tau^{\prime})}
{\partial\,\tau}\right)^{2}+\,
\frac{1}{2}\,v^{''}\,(q_{x_{i}}\,(\tau))\,R^{2}_{x_{i}}\,(\tau,\,\tau^{\prime})
\!\right)>\,0
\label{6.21}
\end{equation}
for $v^{''}\,(q_{x_{i}})\,>\,0$. Eq. (\ref{6.21}) 
follows from (\ref{6.12}) multiplied
by $R_{x_{i}}\,(\tau,\,\tau^{\prime})$ and integrated over $\tau$.

Finally, in our {\bf third approach} we go back to eq. (\ref{4.5}). 
The two matrix elements of the statistical operator
under the $\beta^{\prime}$-integral can be expressed as path integrals
separately. Stationarity of
\renewcommand{\theequation}{7.22} 
\begin{equation}
\begin{array}{c}
\displaystyle{
-\int\limits_{0}^{\beta} d\beta^{\prime}\,\braket{x_{i}|\exp\,(-(\beta\,-\,
\beta^{\prime})\,H)}{x^{\prime}}\,
\braket{x^{\prime}|\exp\,(-\beta^{\prime}\,H)}{x_{i}}\,=}\\
\mbox{}\\
\displaystyle{
-\int\limits_{0}^{\beta}\!d\beta^{\prime}\!\!\!\int\limits_{q_{2}\,(\beta^{\prime}
\,\hbar)\,=\,x^{\prime}}^{q_{2}\,(\beta\,\hbar)\,=\,x_{i}}\!\!\!\!\!
Dq_{2}\,(\tau)\,\exp\,\left\{\!-\frac{1}{\hbar}\,S_{2}\,[q_{2}]\right\}\!
\!\!\int\limits_{q_{1}\,(0)\,=\,x_{i}}^{q_{1}\,(\beta^{\prime}\,\hbar)\,
=\,x^{\prime}}\!\!\!\!\!\!
Dq_{1}\,(\tau)\,\exp\,\left\{\!-\frac{1}{\hbar}\,S_{1}\,[q_{1}]\right\},}
\end{array}
\label{6.22}
\end{equation}
with respect to paths $q_{1}\,(\tau)$ and $q_{2}\,(\tau)$ for $S_{1}
\,[q_{1}]$ and $S_{2}\,[q_{2}]$ defined in (\ref{4.6}), 
leads to the usual equations of motion
and boundary conditions as indicated above. Stationarity with respect to
$\beta^{\prime}$ results in
\renewcommand{\theequation}{7.23} 
\begin{equation}
0=\frac{\partial}{\partial\,\beta^{\prime}}\,(S_{1}\,[q_{1}] + S_{2}\,[q_{2}])=
\frac{m}{2}\,\dot{q}^{2}_{1}\,(\beta^{\prime}\,\hbar)+v\,(q_{1}\,(\beta^{\prime}\,
\hbar))-
\frac{m}{2}\,\dot{q}^{2}_{2}\,(\beta^{\prime}\,\hbar)-v\,(q_{2}\,(\beta^{\prime}\,
\hbar))\,.
\label{6.23}
\end{equation}
Since $v\,(q_{1}\,(\beta^{\prime}\,\hbar))\,=\,v\,(q_{2}\,(\beta^{\prime}\,\hbar))$
by virtue of the boundary conditions, we have
$\dot{q}_{1}\,(\beta^{\prime}\,\hbar)\,=\,\pm\,\dot{q}_{2}\,(\beta^{\prime}\,\hbar)$
for the velocities, and the energies $E_{j}\,=\,
\frac{1}{2}\,m\,\dot{q}^{2}_{j}\,-\,v\,(q_{j})$ are the same for both paths
with $j\,=\,1$ and $j\,=\,2$. Our final result is
\renewcommand{\theequation}{7.24} 
\begin{equation}
\frac{
\frac{\delta}{\delta\,v\,(x^{\prime})}\,\braket{x_{i}|\E^{-\beta\,H}}{x_{i}}}
{\braket{x_{i}|\E^{-\beta\,H}}{x_{i}}}\,
=\,
-\frac{A_\beta A_{1}\,A_{2}}{A_{x_{i}}}
\,\exp\,
\left\{-\frac{1}{\hbar}\,(S_{1}\,[q_{1}]\,+\,S_{2}\,[q_{2}]\,
-\,S\,[q_{x_{i}}])\right\}
\label{6.24}
\end{equation}
where $q_{x_{i}}$ is the solution of (\ref{5.2}) 
with $q_{x_{i}}\,(0)\,=\,x_{i}\,
=\,q_{x_{i}}\,(\beta\,\hbar)$, $A_\beta$ is a norming constant 
due to the stationary phase approximation of the $\beta^\prime$--integral
in (\ref{6.22})
and $A_{1}$, $A_{2}$, 
$A_{x_{i}}$ stand for the fluctuations around the classical solutions as in
(\ref{5.7}), (\ref{5.9}). 

\renewcommand{\thesection}{8}
\section{Numerical case study}
In this section we present numerical results for a simple, one-dimensional
model, which merely serve
to demonstrate that the path integral technique can be used in actual
practice within the Bayesian
approach to inverse quantum statistics. We will discuss in turn the
classical equations of
motion (\ref{310}) with boundary conditions (\ref{311}) and the stationarity
equations (\ref{313}) of the maximum posterior
approximation, which eventually have to be solved simultaneously.

For a numerical implementation we discretize both the time $\tau$,
parametrizing some classical path
$q\,(\tau)$, and the position coordinate $x$, upon which the potential
$v\,(x)$ depends. The time interval $[0,\,\beta\,\hbar]$ is divided into
$n_{\tau}$ equal  steps of length
\renewcommand{\theequation}{8.1} 
\begin{equation}
\varepsilon\,=\,\beta\,/\,n_{\tau}
\label{7.1}
\end{equation}
choosing units such that $\hbar\,=\,1$. A path $q\,(\tau)$ is then coded as
vector $\vec{q}$ with components $q_{k}\,=\,q\,(\tau_{k})$ for
$\tau_{k}\,=\,\varepsilon\,k$; $k\,=\,0,\,1,\,...\,,\,n_{\tau}$. The
potential $v\,(x)$ is studied on an equidistant mesh of size $n_{x}$ in
space, choosing
$n_{x}\,=\,n_{\tau}$. To match the equidistant values of coordinate $x$ to
the corresponding values of the classical path
$q\,(\tau)$ we may either round up or down the function values $q_{k}$ or
linearly interpolate the potential between
equidistant $x$-values.

In their discretized version, the classical equations of motion of our
fictitious particle in
potential $-v\,(q)$ read
\renewcommand{\theequation}{8.2} 
\begin{equation}
0\,=\,-\frac{m}{\varepsilon^{2}}\,(q_{k\,+\,1}\,-\,2\,q_{k}\,+\,q_{k\,
-\,1})\,+\,
v^{\prime}\,(q_{k})\,\,; \qquad k\,=\,1,\,2,\,...\,n_{\tau\,-\,1}\,\,,
\label{7.2}
\end{equation}
and are to be solved with boundary conditions
\renewcommand{\theequation}{8.3} 
\begin{equation}
q_{0}\,=\,x\,=\,q_{n_{\tau}}\,\,.
\label{7.3}
\end{equation}
Eqs. (\ref{7.2}) and (\ref{7.3}) 
amount to solving the matrix equation, for given $v\,(q)$,
\renewcommand{\theequation}{8.4} 
\begin{eqnarray}
0
& = &
-\frac{m}{\varepsilon^{2}}\,\left(\begin{array}{ccccccc}
\frac{\varepsilon^{2}}{m} & 0 & \cdots & & & \cdots & 0\\
1 & -2 & 1 & 0 & \cdots & & 0\\
0 & 1 & -2 & 1 & 0 & \cdots  & \vdots\\
\vdots & & & & & & \\
0 & \cdots & & & 1 & -2 & 1\\
0 & \cdots & & & \cdots & 0 & \frac{\varepsilon^{2}}{m}
\end{array}\right) \left( \begin{array}{c}
q_{0}\\ q_{1}\\ \vdots\\ \mbox{}\\ q_{n_{\tau}\,-\,1}\\
q_{n_{\tau}} \end{array}\right) 
\nonumber\\&&
+
\left( \begin{array}{c}
x\\ v^{\prime}\,(q_{1})\\
\vdots\\ \mbox{}\\
v^{\prime}\,(q_{n_{\tau}\,-\,1})\\ x\end{array} \right)\nonumber\\
\mbox{}\nonumber\\
& \equiv & A\,\vec{q}\,+\,\vec{t}\,(\vec{q}\,)
\label{7.4}
\end{eqnarray}
which is done by iteration according to
\renewcommand{\theequation}{8.5} 
\begin{equation}
\vec{q}^{\,(j\,+\,1)}\,=\,\vec{q}^{\,(j)}\,-\,\eta_{q}\,\left(\vec{q}
^{\,(j)}\,+\,A^{-1}\,\vec{t}\,
\left(\vec{q}^{\,(j)}\right)\right)\,\,.
\label{7.5}
\end{equation}
Step length $\eta_{q}$ in (\ref{7.5}) can be adapted during iteration. Having
solved (\ref{7.4}) for various boundary values $x_{i}$, we can calculate the
likelihood
$p\,(x_{i}|O_{i},\,V)$, eq. (2.14), in classical (eq. (\ref{5.18})) and
semiclassical (eq. (\ref{315})) approximations, and the exact quantum
statistical result (see appendix).

As example we consider a potential of the form
\renewcommand{\theequation}{8.6} 
\begin{equation}
v\,(x)\,=\,-\frac{1}{1\,+\,\exp\,\left(\frac{1}{2}\,(|x\,-\,15|-4)\right)}
\label{7.6}
\end{equation}
on an equidistant mesh with $n_{x}\,=\,n_{\tau}\,=\,30$, shown in Fig.~3.,
upper left part. The right hand
side of Fig.~3 displays the potential $-v\,(q)$ together with the range and
energy of solutions $q_{x}\,(\tau)$ of (\ref{7.4})
for various boundary values $x$ (upper part), and the solutions $q_{x}\,
(\tau)$ as functions of $\tau$.
Note that solutions $q_{x}\,(\tau)$ refer to a boundary value problem in
the fictitious potential $-v\,(q)$ rather than to the
initial value problem of classical mechanics in potential $+v\,(q)$. The
probabilities
$p\,(x_{i}|O_{i},\,V)$, eq. (2.14), in the lower left part of Fig.~3
exhibit the difference
of the classical and semiclassical approximations to the exact quantum
statistical result.
As expected on account of the uncertainty relation,
the variance of the probability distribution increases when going from the
classical limit to the exact quantum mechanical
calculation. The 3 curves coincide in the classical result, if temperature
or mass are increased.

To evaluate the first term of stationarity equation (\ref{313}) for
one-dimensional models,
one should not use eq. (\ref{314}): 
on every one-dimensional, periodic path the
velocity takes the
value zero for at least one value of $\tau$. A zero of the argument of the
$\delta$-distribution at that
value of $\tau$ will not be a simple one. We have, therefore, for the
discretized values $\tau_{k}$, replaced the value
$q\,(\tau_{k})$ by the nearest integer of the interval $[0,\,n_{x}]$, and
the
$\delta$-distribution in (\ref{314}) by the Kronecker symbol, hence
\renewcommand{\theequation}{8.7} 
\begin{equation}
\int d\tau\,\delta\,(q_{i}\,(\tau)\,-\,x)\quad \to \quad
\varepsilon \sum\limits_{j\,=\,1}^{n_{\tau}} \delta_{q_{ij}\,x}\,\,.
\label{7.7}
\end{equation}
The matrix elements of the second term $\braket{x|\exp\,(-\beta\,H)}{x}$
are calculated semiclassically according
to (\ref{315}). In the prior (2.10) we use
\renewcommand{\theequation}{8.8} 
\begin{equation}
\Gamma\,[v]\,=\,-\sum\limits_{i,\,j}\,v_{i}\,\Delta_{ij}\,v_{j}
\label{7.8}
\end{equation}
with
\renewcommand{\theequation}{8.9} 
\begin{equation}
\Delta_{ij}\,=\,\frac{1}{\varepsilon^{2}}\,\left( \begin{array}{cccccccc}
-2 & 1 & 0 & \cdots & & & & 0\\
1 & -2 & 1 & & & & & \vdots\\
\vdots & & \cdots & & & & & \\
 & & & \cdots & & & & \\
 & & & & & 1 & -2 & 1\\
0 & & & \cdots & & 0 & 1 & -2
\end{array} \right)\,\,,
\label{7.9}
\end{equation}
thus demanding smoothness for the potential $v\,(x)$ to be determined. In our
actual calculation we have
sampled $N\,=\,15$ data from the discretized version $(n_{x}\,=\,30$) of
potential
\renewcommand{\theequation}{8.10} 
\begin{equation}
v\,(x)\,=\,\left\{\begin{array}{cc}
\displaystyle{
\frac{1}{4}\,\left(\cos\,\left(\frac{2\,\pi}{10}\,(x\,-\,15)\right)\,
-\,1\right)} & \text{for}\,\,\,
x\,\in\,[5,25]\\
\mbox{}\\
0 & \text{elsewhere} \end{array} \right.
\label{7.10}
\end{equation}
for $\beta\,=\,10$. Eq. (\ref{313}) is then solved, simultaneously with eqs.
(\ref{7.4}), (\ref{7.5}), by iteration, using
the gradient descent algorithm. The hyperparameter $\gamma$ is chosen such
that the depths of the reconstructed potential
and the true potential (\ref{7.10}) 
are approximately equal as shown on the left
hand side of Fig.~4. The right hand
side shows the empirical density of data together with the likelihood for
the true potential and for the classical, semiclassical and quantum
mechanical reconstructions. For sufficiently heavy masses the gross shape
of the
potential is recognized; classical, semiclassical and quantum mechanical
likelihoods are approximately
the same. With decreasing mass, the differences of classical, semiclassical
and quantum
mechanical likelihoods become more pronounced, with the double--hump
structure of the potential
still recognized (Fig.~5). To better reproduce the absolute value of the
potential
minima one may decrease parameter $\gamma$ at the expense of distorting the
symmetrical shape of the potential,
like in Fig.~4.

\renewcommand{\thesection}{9}
\section{Conclusion}
In this paper we have developed the inverse problem of quantum statistics
in path integral representation which supplements the energy representation
used in a number of recent publications.
The advantage of the path integral representation in this context
turns out to be twofold:
First, one can study the semiclassical and classical limits which are
of interest for the analysis of experimental data as obtained from 
atomic force microscopy.
Second, with the path integral representation for the likelihood and the
functional integration over possible potential fields,
one obtains a unified description for the basic equations of
Bayesian inverse quantum statistics. Various approximation schemes
have been studied for calculating, in this representation, 
the statistical operator and its derivatives which are the essential quantities
in maximum posterior approximation.
In particular, the classical limit is obtained and quadratic quantum 
fluctuations are calculated. A simple numerical example is presented
to demonstrate the actual applicability of this approach
which is expected to be useful for analysing experimental data
when spatial distances are reached which resolve nanostructures 
like in recent atomic force microscopy.

\renewcommand{\thesection}{\mbox{}}
\section{Appendix}
 Matrix elements $\braket{x^{\prime}|\E^{-\beta\,H}}{x}$ of the statistical
 operator and their functional derivatives with
respect to $V$, needed in eq. (2.19), are easily calculated in energy
representation. We
start from the Schr\"odinger equation
\renewcommand{\theequation}{A.1}
\begin{equation}
H\,\ket{\phi_{\alpha}}\,=\,(T\,+\,V)\,\ket{\phi_{\alpha}}\,=\,E_{\alpha}\,
\ket{\phi_{\alpha}}\,\,,
\end{equation}
together with orthonormality and closure of eigenfunctions,
\renewcommand{\theequation}{A.2}
\begin{equation}
\braket{\phi_{\alpha^{\prime}}}{\phi_{\alpha}}\,=\,\delta_{\alpha^{\prime}\alpha}\,
\,,\qquad
\sum\limits_{\alpha}\,\ket{\phi_{\alpha}}\,\bra{\phi_{\alpha}}\,=\,{\rm
1\!l}\,\,.
\end{equation}
For the derivatives of
\renewcommand{\theequation}{A.3}
\begin{eqnarray}
\braket{x^{\prime}|\E^{-\beta\,H}}{x}
& = &
\sum\limits_{\alpha}\,\braket{x^{\prime}}{\phi_{\alpha}}\,
\E^{-\beta\,E_{\alpha}}\,\braket{\phi_{\alpha}}{x}\nonumber\\
& = &
\sum\limits_{\alpha}\,\phi^{*}_{\alpha}\,(x)\,\phi_{\alpha}\,(x^{\prime})\,
\E^{-\beta\,E_{\alpha}}
\end{eqnarray}
and of
\renewcommand{\theequation}{A.4}
\begin{eqnarray}
Z
& = &
\int dx\,\braket{x|\E^{-\beta\,H}}{x}\,= \int dx\sum\limits_{\alpha}\,
\braket{x}{\phi_{\alpha}}\,\E^{-\beta\,E_{\alpha}}\,\braket
{\phi_{\alpha}}{x}\nonumber\\
& = &
\sum\limits_{\alpha} \int dx\,\phi^{*}_{\alpha}
\,(x)\,\phi_{\alpha}\,(x) \,\E^{-\beta\,E_{\alpha}}\,=\,
\sum\limits_{\alpha} \E^{-\beta\,E_{\alpha}}
\end{eqnarray}
we need $\delta\,E_{\alpha}\,/\,\delta\,v\,(x)$ and $\delta\,\phi_{\alpha}
\,(x^{\prime})\,/\,
\delta\,v\,(x)$. These derivatives are obtained by variation of the
Schr\"odinger equation,
\renewcommand{\theequation}{A.5}
\begin{equation}
\frac{\delta\,H}{\delta\,V}\,\ket{\phi_{\alpha}}\,+\,H\,\ket{\delta_{V}\,
\phi_{\alpha}}\,=\,
\frac{\delta\,E_{\alpha}}{\delta\,V}\,\ket{\phi_{\alpha}}\,+\,E_{\alpha}\,
\ket{\delta_{V}\,\phi_{\alpha}}
\end{equation}
where, for a local potential 
$V\,(x^{\prime},\,x^{''})\,=\,v\,(x^{''})\,\delta\,
(x^{\prime}\,-\,x^{''})$,
\renewcommand{\theequation}{A.6}
\begin{equation}
\braket{x^{\prime}|\frac{\delta\,H}{\delta\,v\,(x)}}{x^{''}}\,=\,
\frac{\delta\,H\,(x^{\prime},\,x^{''})}{\delta\,v\,(x)}\,=\,
\delta\,(x^{\prime}\,-\,x^{''})\,\delta\,(x\,-\,x^{''})\,\,,
\end{equation}
in short
\renewcommand{\theequation}{A.7}
\begin{equation}
\frac{\delta\,H}{\delta\,v\,(x)}
= \ket{x}\,\bra{x}\,\,.
\end{equation}
\vspace*{1mm}

Multiplying (A.5) by $\bra{\phi_{\alpha}}$ from the left and using the
adjoint of (A.1),
\renewcommand{\theequation}{A.8}
\begin{equation}
\bra{\phi_{\alpha}}\,H\,=\,\bra{\phi_{\alpha}}\,E_{\alpha}\,\,,
\end{equation}
yields with normalization (A.2)
\renewcommand{\theequation}{A.9}
\begin{eqnarray}
\frac{\delta\,E_{\alpha}}{\delta\,v\,(x)}
& = &
\braket{\phi_{\alpha}|\frac{\delta\,H}{\delta\,V}}{\phi_{\alpha}}\nonumber\\
& = &
\braket{\phi_{\alpha}}{x}\,\braket{x}{\phi_{\alpha}}\,=\,|\phi_{\alpha}
\,(x)|^{2}\,\,.
\end{eqnarray}
Hence our first result, in agreement with (\ref{4.3}), is
\renewcommand{\theequation}{A.10}
\begin{eqnarray}
\frac{\delta\,Z}{\delta\,v\,(x)}
& = &
-\beta\,\sum\limits_{\alpha}\,\E^{-\beta\,E_{\alpha}}\,
\frac{\delta\,E_{\alpha}}{\delta\,v\,(x)}
=
-\beta\sum\limits_{\alpha}\,\E^{-\beta\,E_{\alpha}}\,|\phi_{\alpha}\,(x)
|^{2}\nonumber\\
& = &
 -\beta\,\sum\limits_{\alpha} \braket{x}{\phi_{\alpha}}\,
 \E^{-\beta\,E_{\alpha}}
\,\braket{\phi_{\alpha}}{x}\,=\,-\beta\,\braket{x|\E^{-\beta\,H}}{x}\,\,,
\label{A.10}
\end{eqnarray}
with closure (A.2).

To find the derivative of $\phi_{\alpha}\,(x^{\prime})$ with respect to $v\,(x)
$ we rewrite (A.5) as
inhomogeneous equation,
\renewcommand{\theequation}{A.11}
\begin{equation}
(E_{\alpha}\,-\,H)\,\ket{\delta_{V}\,\phi_{\alpha}}
= \left(
\frac{\delta\,H}{\delta\,V}
- \frac{\delta\,E_{\alpha}}{\delta\,V}\right)
\ket{\phi_{\alpha}}\,\,.
\end{equation}
Obviously all orbitals $\ket{\phi_{\alpha^{\prime}}}$ with $E_{\alpha^{\prime}}\,
=\,E_{\alpha}$ are in the
null space of the operator $(E_{\alpha}\,-\,H)$, hence $(E_{\alpha}\,-\,H)
$ is not
invertible in full Hilbert space. However, the Moore-Penrose method of the
pseudo-inverse
can be applied to solve (A.11) for $\ket{\delta_{V}\,\phi_{\alpha}}$. The
solvability condition states that the right
hand side of (A.11) has no component in the null space of $(E_{\alpha}\,
-\,H$). This is easily verified,
multiplying (A.11) by $\bra{\phi_{\alpha^{\prime}}}$ with $E_{\alpha^{\prime}}\,
=\,E_{\alpha}$:
\renewcommand{\theequation}{A.12}
\begin{equation}
\braket{\phi_{\alpha^{\prime}}|E_{\alpha}\,-\,H}{\delta_{V}\,\phi_{\alpha}}\,
=\,0\,=\,
\braket{\phi_{\alpha^{\prime}}|
  \frac{\delta\,H}{\delta\,V}-\frac{\delta\,E_{\alpha}}{\delta\,V}}
{\phi_{\alpha}}\,\,,
\end{equation}
applying $H$ to the left according to (A.8). It is easy to control that
\renewcommand{\theequation}{A.13}
\begin{equation}
G_{\alpha}\,=\,\sum\limits_{\gamma\,\not=\,\alpha}\,\frac{\ket
{\phi_{\gamma}}\bra{\phi_{\gamma}}}{E_{\alpha}\,-\,E_{\gamma}}
\end{equation}
is the pseudo-inverse of $(E_{\alpha}\,-\,H$), fulfilling the condition
\renewcommand{\theequation}{A.14}
\begin{equation}
G_{\alpha}\,(E_{\alpha}\,-\,H)\,G_{\alpha}\,=\,G_{\alpha}\,\,.
\end{equation}
To obtain a unique solution of (A.11), or equivalently (A.5), we demand
that $\ket{\delta_{V}\,\phi_{\alpha}}$ has no
component in the null space of $(E_{\alpha}\,-\,H)$,
\renewcommand{\theequation}{A.15}
\begin{equation}
\braket{\phi_{\alpha^{\prime}}}{\delta_{V}\,\phi_{\alpha}}\,=\,0\quad \text{for
all} \quad
\phi_{\alpha^{\prime}} \quad \text{with}\quad E_{\alpha^{\prime}}\,=\,E_{\alpha}\,\,.
\end{equation}
This corresponds to fixing norm and phase of the eigenstates $\ket
{\phi_{\alpha}}$ and, in case of
degeneracy, uses the freedom to work with arbitrary orthonormal linear
combinations of the respective eigenstates. Applying
the pseudo-inverse $G_{\alpha}$ to (A.11) and using the orthonormality
(A.2) we thus find in the subspace where
$(E_{\alpha}\,-\,H$) is invertible
\renewcommand{\theequation}{A.16}
\begin{equation}
\ket{\delta_{V}\,\phi_{\alpha}}\,=\,\sum\limits_{\begin{array}{c}
\scriptstyle{\gamma}\\
\scriptstyle{
E_{\gamma}\,\not=\,E_{\alpha}}\end{array}} (E_{\alpha}\,-\,E_{\gamma})^{-1}
\,\ket{\phi_{\gamma}}\,
\braket{\phi_{\gamma}|\delta_{V}\,H}{\phi_{\alpha}}\,\,,
\end{equation}
or explicitly in coordinate representation with eq. (A.6)
\renewcommand{\theequation}{A.17}
\begin{equation}
\frac{\delta\,\phi_{\alpha}\,(x^{\prime})}{\delta\,v\,(x)}
=
\sum\limits_{\begin{array}{c}
\scriptstyle{\gamma}\\ \scriptstyle{E_{\gamma}\,\not=\,E_{\alpha}}
\end{array}}
\frac{1}{E_{\alpha}\,-\,E_{\gamma}}\,\phi_{\gamma}\,(x^{\prime})\,\phi_{\gamma}
^{*}\,(x)\,\phi_{\alpha}\,(x)\,\,.
\end{equation}
With the derivatives (A.9) and (A.17), given in terms of the solutions of
the Schr\"odinger equation, we can now construct the functional
derivative of $\braket{x^{\prime}|\E^{\beta\,H}}{x}$ with respect to
$v\,(x^{''})$ according to (A.3):
\begin{equation*}
\frac{\delta}{\delta\,v\,(x^{''})}\,\braket{x^{\prime}|\E^{-\beta\,H}}{x}=
-\beta\,\E^{-\beta\,E_{\alpha}}\sum\limits_{\alpha}\,\phi_{\alpha}^{*}(x)\,
\phi_{\alpha}(x^{\prime})\,|\phi_{\alpha}\,(x^{''})|^{2}
\end{equation*}
\renewcommand{\theequation}{A.18}
\begin{equation}
\end{equation}
\begin{equation*}
+\!\!\!\sum\limits_{\begin{array}{c}\scriptstyle{\alpha,\,\gamma}\\
\scriptstyle{\alpha\,\not=\,\gamma}\end{array}} \frac{\exp
(-\beta\,E_{\alpha})}{E_{\alpha}-E_{\gamma}}\left\{
\phi_{\alpha}^{*}(x)\,\phi_{\gamma}(x^{\prime})\,\phi^{*}_{\gamma}(x^{''})\,
\phi_{\alpha}(x^{''}) +
\phi_{\gamma}^{*}(x)\,\phi_{\gamma}(x^{''})\,\phi_{\alpha}^{*}(x^{''})\,
\phi_{\alpha}(x^{\prime})\right\}.
\end{equation*}
In particular, for diagonal elements with $x^{\prime}\,=\,x$,
\renewcommand{\theequation}{A.19}
\begin{equation}
\begin{array}{c}
\displaystyle{
\frac{\delta}{\delta\,v\,(x^{''})}\,\braket{x|\E^{-\beta\,H}}{x}\,=\,
-\beta\,\E^{-\beta\,E_{\alpha}}\,\sum\limits_{\alpha}\,|\phi_{\alpha}\,(x)
|^{2}\,|\phi_{\alpha}\,(x^{''})|^{2}}\\
\mbox{}\\
\displaystyle{
+\!\!\sum\limits_{\begin{array}{c}\scriptstyle{\alpha,\,\gamma}\\
\scriptstyle{\alpha\,\not=\,\gamma}\end{array}}
\,\frac{\exp\,(-\beta\,E_{\alpha})}{E_{\alpha}\,-\,E_{\gamma}}\,2\,
\text{Re}\,\left\{
\phi_{\alpha}^{*}\,(x)\,\phi_{\gamma}\,(x)\,\phi_{\alpha}\,(x^{''})\,
\phi_{\gamma}^{*}\,(x^{''})\right\}}
\end{array}
\end{equation}
This result, eq. (A.19), is identical with (\ref{4.5}) as can be shown by
carrying out the $\beta^{\prime}$--integration
in (\ref{4.5}) using the energy representation of the matrix elements under the
integral:
\renewcommand{\theequation}{A.20}
\begin{equation}
\begin{array}{c}
\displaystyle{
\frac{\delta\,\braket{x|\E^{-\beta\,H}}{x}}{\delta\,v\,(x^{''})}\,
=-\int\limits_{0}^{\beta}\!\!d\beta^{\prime}\,
\braket{x|\E^{-(\beta -\beta^{\prime})\,H}}{x^{''}} \braket{x^{''}
|\E^{-\beta^{\prime}\,H}}{x}\,=}\\
\mbox{}\\
\displaystyle{
-\sum\limits_{\alpha,\,\gamma}
\phi_{\alpha}(x)\,\phi^{*}_{\alpha}\,(x^{''})\,\phi_{\gamma}\,(x^{''})\,
\phi_{\gamma}^{*}\,(x)\,\times\,\E^{-\beta\,E_{\alpha}} \int\limits_{0}
^{\beta}\!\! d\beta^{\prime}\,\exp\,
(-\beta^{\prime}\,(E_{\gamma} - E_{\alpha}))\,\,,}
\end{array}
\end{equation}
inserting the closure relation (A.2). With the $\beta^{\prime}$-integration
carried out,
\renewcommand{\theequation}{A.21}
\begin{equation}
\int\limits_{0}^{\beta}\!\! d\beta^{\prime}\,\exp\,(-\beta^{\prime}\,
(E_{\gamma}\,-\,E_{\alpha}))\,=\,
\left\{ \begin{array}{ccc}
\beta & \text{for} & \alpha\,=\,\gamma\,\,,\text{or else}\\
\mbox{}\\
\displaystyle{\frac{1}{E_{\gamma}\,-\,E_{\alpha}}} & - &
\displaystyle{\frac{\exp\,(-\beta\,(E_{\gamma}\,-\,E_{\alpha}))}
                   {E_{\gamma} \, - \, E_{\alpha}}}\,\,,
\end{array} \right.
\end{equation}
one obtains
\renewcommand{\theequation}{A.22}
\begin{equation}
\begin{array}{c}
\displaystyle{
\frac{\delta\,\braket{x|\E^{-\beta\,H}}{x}}{\delta\,v\,(x^{''})}\,=\,
-\beta\,\E^{-\beta\,E_{\alpha}}\,
\sum\limits_{\alpha}\,|\phi_{\alpha}\,(x)|^{2}\,|\phi_{\alpha}\,(x^{''})
|^{2}}\\
\mbox{}\\
\displaystyle{
+\sum\limits_{\begin{array}{c}
\scriptstyle{\alpha,\,\gamma}\\
\scriptstyle{\alpha\,\not=\,\gamma}\end{array}}
\frac{\exp\,(-\beta\,E_{\alpha})}{E_{\alpha}\,-\,E_{\gamma}}\,2\,\text{Re}
\,
\left\{\phi_{\alpha}\,(x)\,\phi_{\alpha}^{*}\,(x^{''})\,\phi_{\gamma}
\,(x^{''})\,\phi^{*}_{\gamma}\,(x)\right\}\,\,,}
\end{array}
 \end{equation}
q. e. d.

\clearpage

\begin{figure}
\epsfig{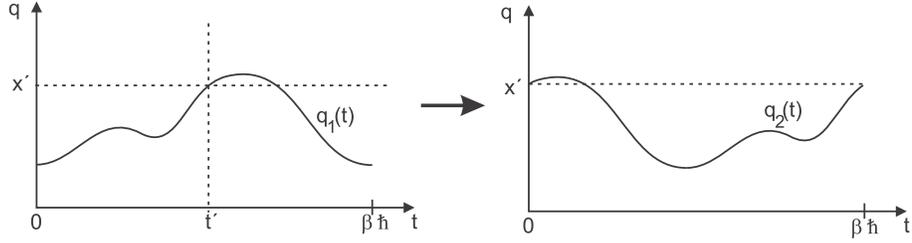}
\caption{
Equivalent paths $q_{1}\,(\tau)$ and $q_{2}\,(\tau)$ in the interval
$[0,\,\hbar\,\beta]$, explaining eq. (\ref{4.2}).
(In the figure $\tau$ is denoted $t$.)}
\end{figure}

\begin{figure}
\begin{center}
\epsfig{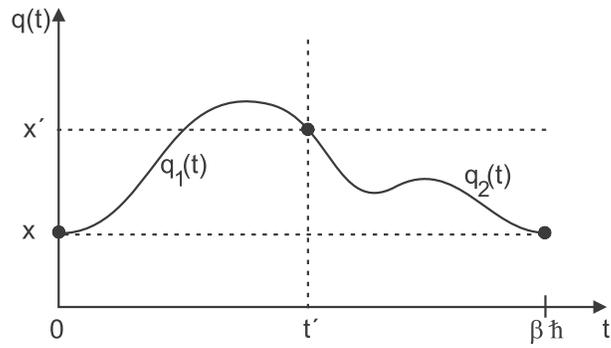}
\end{center}
\caption{
Splitting path $q\,(\tau)$ into parts $q_{1}\,(\tau)$ and $q_{2}\,(\tau)$,
explaining the 
\mbox{transition} 
from eq. (\ref{4.4}) to (\ref{4.5}).
(In the figure $\tau$, $\tau^\prime$ are denoted $t$, $t^\prime$, 
respectively.)}
\end{figure}

\begin{figure}
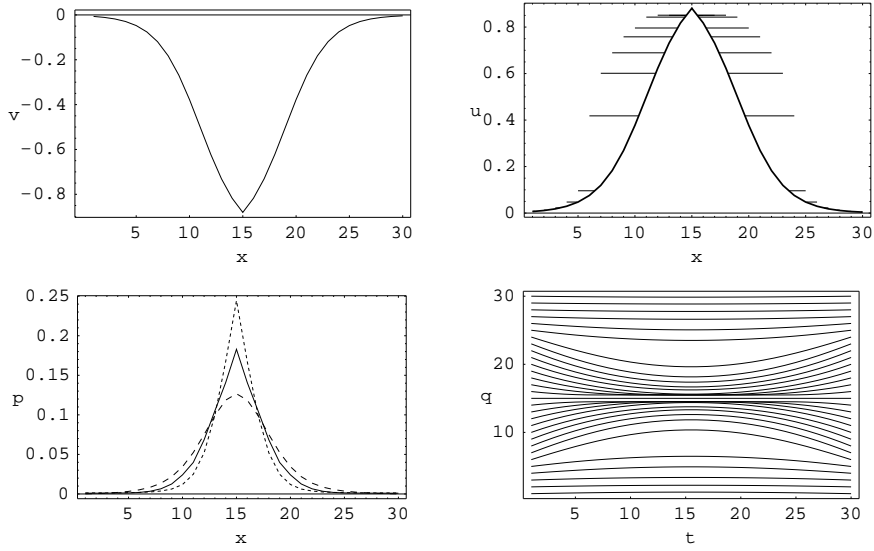

\epsfig{file=ps/d1kap6col.eps,width=60mm}
\epsfig{file=ps/d1kap6cor.eps,width=60mm}
\epsfig{file=ps/d1kap6cul.eps,width=60mm}
\epsfig{file=ps/d1kap6curr.eps,width=60mm}
\caption{ 
Comparison of classical, semiclassical and quantum mechanical likelihood.
Upper left part:
original potential $v\,(x)$. Upper right part: potential $u\,(x)\,=\,
-v\,(x)$, to be used in (\ref{7.2}); thin
horizontal lines indicate range and energy of paths $q_{x}\,(\tau)$. Lower
left part: classical (dotted line), semiclassical
(full line) and quantum mechanical (dashed line) likelihoods. Lower right
part:
paths $q_{x}\,(\tau)$ for various $x$-values. Parameters used are $m\,
=\,0.1$, $\beta\,=\,6$, $n_{\tau}\,=\,30$,
$n_{x}\,=\,30$.
(In the figure $\tau$ is denoted $t$.)
}
\end{figure}

\begin{figure}
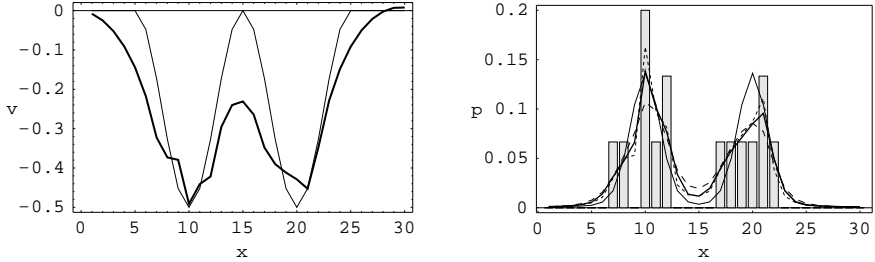

\epsfig{file=ps/d7kap6al.eps,width=60mm}
\epsfig{file=ps/d7kap6ar.eps,width=60mm}
\caption{
Bayesian reconstruction of potentials using the path integral method. Left
part:
original potential (thin line) and reconstructed potential (thick line).
Right part:
relative frequencies of sampled data (bars), likelihood of the true (thin
line) potential
and of the reconstructed potential: semiclassical density obtained by
iteration (thick line), classical
(dotted line) and quantum mechanical density (dashed line). Parameters:
$\beta\,=\,10$, $m\,=\,1$, $\gamma\,=\,5$, $N\,=\,15$.
}
\end{figure}

\begin{figure}
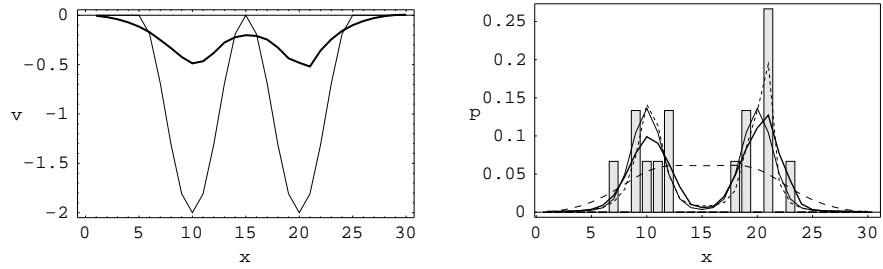

\epsfig{file=ps/d4kap6al.eps,width=60mm}
\epsfig{file=ps/d4kap6ar.eps,width=60mm}
\caption{
Bayesian reconstruction of potentials, using the path integral method, for
small masses: Graphics as in Fig.~4. Parameters:
$\beta\,=\,10$, $m\,=\,0.05$, $\gamma\,=\,10$, $N\,=\,15$.
}
\end{figure}


\begin{thebibliography}{10}
\bibitem{1}
A. N. Tikhonov and V. Arsenin,
{\em Solution of Ill-posed Problems}
(John Wiley, New York, 1977);
\newblock
V. N. Vapnik, 
{\em Statistical Learning Theory}
(John Wiley, New York, 1998);
\newblock
A. Kirsch, 
{\em An Introduction to the Mathematical Theory of Inverse Problems}
(Springer Verlag, New York, 1996).
\bibitem{2}
R. G. Newton,
{\em Inverse Schr\"odinger Scattering in Three Dimensions}
(Springer Verlag, Berlin, 1989)
\bibitem{3}
K. Chadan, D. Colton, L. P\"aiv\"arinta, and W. Rundell,
{\em An Introduction to Inverse Scattering 
and Inverse Spectral Problems}
(SIAM, Philadelphia, 1997);
\newblock
B. N. Zakhariev and V. M. Chabanov,
{\em Inverse Problems} {\bf 13} (1997) R47--R79.
\bibitem{4}
J. C. Lemm,
{\em Bayesian Field Theory}
(The Johns Hopkins University Press, Baltimore, 2003).
\bibitem{5}
J. C. Lemm, J. Uhlig, and A. Weiguny,
{\em Phys. Rev. Lett.} {\bf 84} (2000) 2068;
\newblock
{\em The European Physical Journal B} {\bf 20} (2001) 349;
\newblock
J. C. Lemm and J. Uhlig, 
{\em Few Body Systems} {\bf 29} (2000) 25;
{\em Phys. Rev. Lett.} {\bf 84} (2000) 4517;
\newblock
J. C. Lemm, 
{\em Phys. Lett. A} {\bf 276} (2000) 19.
\bibitem{6}
J. Zinn--Justin,
{\em Quantum Field Theory and Critical Phenomena}
(Clarendon Press, Oxford 1996);
\newblock
L. S. Schulman,
{\em Techniques and Applications of Path Integration}
(John Wiley, New York, 1981);
\newblock
J. W. Negele and H. Orland,
{\em Quantum Many--Particle Systems}
(Addison--Wesley, Redwood City, CA, 1988).
\bibitem{Fuchs}
B. Gotsmann and H. Fuchs,
{\em Phys. Rev. Lett.} {\bf 86} (2001) 1872
\bibitem{7}
H. R\"omer and T. Filk,
{\em Statistische Mechanik}
(VCH, Weinheim, 1994)
\bibitem{8}
R. F. Dashen, B. Hasslacher, and A. Neveu,
{\em Phys. Rev. D} {\bf 10} (1974) 4114;
H. Kleinert, {\em Pfadintegrale}
(Wissenschaftsverlag, Mannheim, 1993)
\end{thebibliography}
\end{document}